\documentclass[aps,pre,reprint,superscriptaddress,footinbib,floatfix]{revtex4-2}

\usepackage{graphicx}
\usepackage{bm}
\usepackage{amssymb}
\usepackage{amsbsy}
\usepackage{amsmath}
\usepackage{mathtools}
\usepackage{xcolor}
\usepackage{stmaryrd}
\usepackage{mathrsfs}
\usepackage{tikz}
\usetikzlibrary{shapes}
\usepackage{cancel}
\usepackage[normalem]{ulem}
\usepackage{hyperref}
\usepackage{multirow}

\usepackage{float}

 \usepackage{placeins}

\makeatletter
\let\cat@comma@active\@empty
\makeatother

\makeatletter 
\gdef\@ptsize{2}
\let\@currsize\normalsize 
\makeatother 

\usepackage{graphicx}
\usepackage{dcolumn}
\usepackage{bm}

\usepackage[utf8]{inputenc}
\usepackage[T1]{fontenc}
\usepackage{mathptmx}
\usepackage{etoolbox}
\usepackage[shortlabels]{enumitem}
\makeatletter
\def\@email#1#2{%
	\endgroup
	\patchcmd{\titleblock@produce}
	{\frontmatter@RRAPformat}
	{\frontmatter@RRAPformat{\produce@RRAP{*#1\href{mailto:#2}{#2}}}\frontmatter@RRAPformat}
	{}{}
}%

\makeatother

\begin{document}


\title[Linear Viscoelasticity of Semidilute Flexible Polymers]{Linear Viscoelasticity of Semidilute Unentangled Flexible Polymer Solutions}
\author{Amit Varakhedkar}
\affiliation{IITB-Monash Research Academy, Indian Institute of Technology Bombay, Mumbai, 400076, India}
\affiliation{Department of Chemical Engineering, Indian Institute of Technology Bombay, Mumbai, 400076, India}
\affiliation{Department of Chemical and Biological Engineering, Monash University, Melbourne, VIC 3800, Australia}

\author{P. Sunthar}
\affiliation{Department of Chemical Engineering, Indian Institute of Technology Bombay, Mumbai, 400076, India}

\author{J. Ravi Prakash}
\email{ravi.jagadeeshan@monash.edu}
\homepage{https://users.monash.edu.au/~rprakash/}
\affiliation{Department of Chemical and Biological Engineering, Monash University, Melbourne, VIC 3800, Australia}

\date{\today}

\begin{abstract}
The linear viscoelastic response of flexible polymer solutions in the dilute and semidilute unentangled regimes is investigated using Brownian dynamics simulations. The relaxation modulus and dynamic moduli are computed over a wide range of concentrations and chain discretizations for both $\theta$ and good solvents. In the dilute limit, simulations recover Zimm-like behavior with solvent-quality-dependent power-law scaling in the intermediate time and frequency regimes, while in the semidilute unentangled regime a systematic crossover to Rouse-like dynamics is observed with increasing concentration due to screening of excluded-volume and hydrodynamic interactions. Comparison with experimental measurements shows excellent agreement for the storage modulus across both concentration regimes and for the loss modulus at low and intermediate frequencies, with deviations at high frequencies due to finite-chain discretization effects. Systematic accounting for these finite-chain length effects using successive fine-graining enables quantitative prediction of the loss modulus in the infinite-chain length limit.
\end{abstract}

\maketitle


\section{\label{intro}Introduction}

To date, extensive experimental studies have investigated the equilibrium and non-equilibrium rheological properties of dilute and semidilute polymer solutions, revealing general physical trends that are largely independent of specific polymer chemistry. Of particular importance among these properties is the linear viscoelastic response, which provides a fundamental link between microscopic polymer dynamics and macroscopic material behavior. Linear viscoelasticity is quantified through the frequency-dependent storage and loss moduli, $G'(\omega)$ and $G''(\omega)$, which probe relaxation processes occurring over a broad range of timescales. These moduli typically exhibit rich frequency dependence, including distinct power-law regimes at intermediate frequencies, arising from the interchain interactions, hydrodynamic interactions, and concentration-dependent screening effects. Understanding how these mechanisms govern the linear viscoelastic response of polymer solutions in the dilute and semidilute regimes is therefore essential for developing a unified description of polymer dynamics across length and time scales.

The concentration regimes of polymer solutions are conventionally characterized using the reduced concentration $c/c^*$, where $c^*$ denotes the overlap concentration at which polymer coils begin to touch each other. In the dilute regime ($c/c^* \ll 1$), polymer chains are effectively isolated and do not experience intermolecular interactions. In contrast, the polymer chains in the semidilute unentangled regime ($c/c^* \geqslant 1$) are characterized by significant chain overlap without the formation of entanglements, which occurs above the entanglement concentration ($c_e$). While reptation and tube-like dynamics are absent in this regime, chain conformations and dynamics are strongly modified by intermolecular interactions, as well as by hydrodynamic screening and excluded-volume effects, leading to qualitatively different viscoelastic behavior compared to the dilute limit \citep{Doi1988,Rubinstein2003}.

In the dilute limit, well-established theoretical models exist to describe the linear viscoelastic response of polymer solutions. The Rouse model \citep{Rouse1953} provides a description for free-draining chains without hydrodynamic interactions, while the Zimm model \citep{Zimm1956} incorporates hydrodynamic interactions in a pre-averaged manner and successfully predicts the linear viscoelastic response of polymers in $\theta$ solvents. These models yield explicit expressions for the relaxation spectrum and dynamic moduli, derived from normal-mode analysis of the chain dynamics \citep{Doi1988}. However, for good solvents, where excluded-volume interactions lead to non-Gaussian chain statistics, no corresponding first-principles theory for the linear viscoelastic response exists. The simultaneous inclusion of excluded-volume and hydrodynamic interactions renders an exact analytical treatment intractable. As a result, theoretical descriptions rely on approximate approaches. In particular, within the Zimm framework, scaling arguments have been used to obtain approximate expressions for the dynamic moduli in good solvents, based on an assumed relaxation spectrum rather than a rigorous normal-mode analysis \citep{Doi1988, Rubinstein2003}. Within this framework, \citet{Tschoegl1964} extended Zimm’s approach by incorporating excluded-volume effects through modified relaxation spectra and proposed an approximate procedure for calculating the dynamic moduli. Subsequent work further examined alternative approximate schemes for accounting for excluded-volume effects within bead-spring models, while retaining Gaussian dynamics \citep{Sahouani1992}. In all the above approaches, hydrodynamic interactions are treated in a pre-averaged manner and fluctuations in hydrodynamic interactions are neglected.

As the monomer concentration increases, polymer chains begin to interact hydrodynamically and through excluded-volume effects until the overlap concentration $c^*$ is reached, at which individual coils start to touch and interpenetrate. For concentrations above $c^*$ in a good solvent, polymer chains transition from self-avoiding walk (SAW) statistics to random walk (RW) statistics due to the screening of excluded-volume interactions beyond the correlation blob length scale, a phenomenon known as Flory screening \citep{deGennes1979,Rubinstein2003}. The screening of excluded-volume interactions in the semidilute regime is accompanied by a corresponding screening of hydrodynamic interactions beyond the same correlation blob length scale, as predicted by scaling arguments \citep{deGennes1979,Doi1988,Prakash2019}. \citet{Ahlrichs2001} demonstrated that hydrodynamic interactions are present at short times and are only screened at times comparable to the Zimm relaxation time of a correlation blob. The screening of hydrodynamic interactions was also observed in the linear viscoelastic response of poly(AN-co-IA) under $\theta$ solvent conditions from experimental observations by \citet{Zhu2012}, showing that at sufficiently high concentrations in the semidilute unentangled regime, the dynamic moduli are well described by Rouse theory. Complementary computational studies, including Monte Carlo simulations based on lattice models, have independently established the occurrence of Flory screening and validated the predicted crossover from SAW to RW statistics with increasing concentration \citep{Paul1991}.

In the semidilute regime, a complete theoretical description of the linear viscoelastic response of polymer solutions is not available. As concentration increases beyond the overlap threshold, polymer dynamics are influenced not only by interchain interactions but also by concentration-dependent screening of hydrodynamic interactions, resulting in a strongly coupled, multiscale problem. As a result, most theoretical treatments in this regime are limited to asymptotic scaling laws for dynamic properties of polymer solutions, rather than mode-resolved analytical theories for the linear viscoelastic response, for both $\theta$ solvents and good solvents \citep{deGennes1979,Doi1988,Rubinstein2003}. A wide range of experimental studies have established the validity of these scaling predictions for dynamic quantities such as the polymer contribution to the zero-shear viscosity, single-chain diffusivity, and the longest relaxation time in semidilute unentangled solutions. These results have been confirmed for $\theta$ solvent conditions in several experimental systems \citep{Adam1984,Zhu2012,Inoue2002,Pan2021}, and for good solvent conditions in corresponding studies \citep{Heo2005,Clasen2006,Chen2018}, providing strong empirical support for the underlying scaling framework.

Frequency-dependent linear viscoelastic measurements in dilute and semidilute polymer solutions have been reported in a limited number of experimental studies, notably by \citet{Johnson1970}, \citet{Clasen2006}, and \citet{Zhu2012}, which provide detailed data for the storage and loss moduli over wide frequency ranges and for different solvent conditions. However, to date, there are no simulation or theoretical studies that quantitatively reproduce these experimental results across the full frequency spectrum. Incorporating fluctuating hydrodynamic interactions presents a significant challenge within computational frameworks that are traditionally employed to study static properties of polymer solutions. \citet{Huang2010} employed multiparticle collision dynamics (MPCD) coupled with molecular dynamics, providing a computational framework that explicitly incorporates fluctuating hydrodynamic interactions. Although their study did not address the linear viscoelastic response, it validated scaling predictions for dynamic quantities such as the longest relaxation time and the polymer contribution to viscosity in a perfectly good solvent, demonstrating consistency with semidilute unentangled scaling laws.

Previous work has demonstrated that bead-spring Brownian dynamics simulations incorporating hydrodynamic interactions implicitly via the Rotne-Prager-Yamakawa tensor accurately capture the linear viscoelastic response of polymers in the infinitely dilute limit across a range of chain flexibilities \citep{Varakhedkar2025}. Extending this framework, the present study investigates the linear viscoelastic response of flexible polymer solutions in both dilute and semidilute unentangled regimes under $\theta$ and good-solvent conditions, with particular emphasis on the screening of excluded-volume and hydrodynamic interactions at elevated concentrations and on a direct comparison with available experimental measurements.

A further challenge in comparing simulations with experiments arises from the use of finite chain discretization. In simulations, polymer chains are represented using a finite number of beads in bead-spring chain models, which leads to truncation of the relaxation spectrum at high frequencies. To accurately capture the frequency-dependent linear viscoelastic response at higher frequencies and enable direct comparison with experimental data, it is therefore essential to account for finite-chain length effects and systematically approach the infinite-chain length limit. This is achieved through the successive fine-graining procedure \citep{Pham2008,Sunthar2005,Prabhakar2004,Sasmal2017}, wherein a bead-spring chain model is employed and simulation data for finite chains is extrapolated to the long chain length limit or to the number of Kuhn steps in the chain depending on the context. Successive fine graining has proven to be a powerful methodology for assessing whether experimental measurements can be quantitatively reproduced by simulation predictions. The approach has been successfully applied to predict both equilibrium and nonequilibrium properties, including single-chain diffusivity \citep{Jain2012} and uniaxial elongational viscosity \citep{Prabhakar2004,Sunthar2005,Saadat2015}. Extrapolation to long chain length limit has also been employed to obtain universal predictions in shear flows \citep{Ottinger1987,Ottinger1989,Prakash1997,Prakash2002,Kroger2000}. In the present study, this methodology is employed to predict the linear viscoelastic response of polymer chains and to enable quantitative comparison with experimental data in the high-frequency regime.

The remainder of this paper is organized as follows. The bead-spring chain model for flexible polymers and the governing equations, including excluded-volume and hydrodynamic interactions, are introduced first. This is followed by a description of the simulation methodology and parameter choices, along with the definitions of the linear viscoelastic quantities evaluated in this work. The linear viscoelastic response in the infinitely dilute limit is then presented, where the stress relaxation modulus and dynamic moduli are examined under $\theta$ and good solvent conditions and compared with theoretical predictions. The analysis is subsequently extended to semidilute unentangled solutions, highlighting the concentration-induced crossover from Zimm-like to Rouse-like dynamics arising from hydrodynamic screening. Direct comparison between simulation predictions and experimental measurements in the dilute limit is then discussed, where successive fine-graining is employed to account for finite-chain discretization effects and extrapolate viscoelastic quantities to the infinite-chain length limit. This is followed by comparison with experimental data in the semidilute unentangled regime. Finally, the main findings are summarized and directions for future work are outlined.

\section{\label{sec:methods} Model formulation}

\subsection{\label{sec:gov_eqn} Governing equations}

A coarse-grained bead-spring chain model is considered to simulate solutions of flexible linear polymers using Brownian dynamics. The position vector $\bm{r}_{\mu}(t)$ of each bead $\mu$ is evolved in time $t$ using a first-order Euler integration scheme, which numerically solves the It\^o stochastic differential equation governing its Brownian motion \citep{Ottinger2012}:
\begin{align}\label{eq:bd}
	\bm{r}_{\mu}(t + \Delta t) =\;& \bm{r}_{\mu}(t) 
	+ \frac{\Delta t}{4} \sum_{\nu=1}^{N} \bm{D}_{\mu \nu} \cdot \bm{F}_{\nu}
	+ \frac{1}{\sqrt{2}} \sum_{\nu=1}^{N} \bm{B}_{\mu \nu} \cdot \Delta \bm{W}_{\nu}.
\end{align}
The governing equations are non-dimensionalised using Hookean (Brownian dynamics) units with the characteristic length scale $l_H=\sqrt{k_BT/\hat{H}}$, time scale $\lambda_H=\zeta/(4\hat{H})$, force scale $f_H=\sqrt{k_BT\,\hat{H}}$, and energy scale $\epsilon_H=k_BT$, where $\hat{H}$ is the spring stiffness and $\zeta=6\pi\eta_s a$ is the Stokes friction coefficient of a bead of radius $a$ in a solvent of viscosity $\eta_s$. The system is contained within a cubic simulation box of side length $L$ with periodic boundary conditions applied in all directions, such that the total simulation volume is $V = L^3$. The total monomer concentration in the simulation box is defined as $c = N/V$, where $N$ denotes the total number of monomers present in the system. The total number of monomers is given by $N = N_b N_c$, with $N_b$ representing the number of beads per polymer chain and $N_c$ the total number of polymer chains in the simulation box.

The quantity $\Delta \bm{W}_{\nu}$ is a non-dimensional Wiener increment accounting for thermal fluctuations. The components of $\Delta \bm{W}_{\nu}$ are drawn from a Gaussian distribution with zero mean and variance $\Delta t$.
The tensor $\bm{B}_{\mu \nu}$ is obtained from the decomposition of the diffusion tensor $\bm{D}_{\mu \nu}$,
\begin{equation}
	\label{eq:7_3}
	\bm D_{\mu\nu} = \delta_{\mu\nu} \bm{\delta} + \bm{\Omega}_{\mu\nu},
\end{equation}
where $\delta_{\mu \nu}$ is the Kronecker delta, $\bm{\delta}$ is the unit tensor, and $\bm{\Omega}_{\mu \nu}$ is the hydrodynamic interaction tensor.
The matrices $\mathcal{D}$ and $\mathcal{B}$ are block matrices of dimension $3N \times 3N$, such that the $(\mu,\nu)$-th block of $\mathcal{D}$ contains $\bm{D}_{\mu\nu}$, and $\mathcal{B}$ satisfies the decomposition
\begin{equation}
	\mathcal{B}\,\mathcal{B}^{T} = \mathcal{D}.
\end{equation}
Hydrodynamic interactions are modeled using the Rotne-Prager-Yamakawa (RPY) tensor:
\begin{equation}
	\label{eq:7_9}
	\bm{\Omega}_{\mu \nu} = \bm{\Omega}(\bm{r}_{\mu} - \bm{r}_{\nu}),
\end{equation}
with
\begin{equation}
	\bm{\Omega}(\bm{r}) = \Omega_{1} \bm{\delta} + \Omega_{2} \frac{\bm{r}\bm{r}}{r^{2}}.
\end{equation}
The scalar functions $\Omega_{1}$ and $\Omega_{2}$ are given by
\begin{equation*}
	\Omega_1 =
	\begin{cases}
		\dfrac{3\sqrt{\pi}}{4} \dfrac{h^*}{r}\!\left(1+\dfrac{2\pi}{3}\dfrac{{h^*}^2}{r^2}\right),
		& r \ge 2\sqrt{\pi}h^*, \\[8pt]
		1 - \dfrac{9}{32}\dfrac{r}{h^*\sqrt{\pi}},
		& r \le 2\sqrt{\pi}h^*,
	\end{cases}
\end{equation*}

\begin{equation*}
	\Omega_2 =
	\begin{cases}
		\dfrac{3\sqrt{\pi}}{4} \dfrac{h^*}{r}\!\left(1-\dfrac{2\pi}{3}\dfrac{{h^*}^2}{r^2}\right),
		& r \ge 2\sqrt{\pi}h^*, \\[8pt]
		\dfrac{3}{32}\dfrac{r}{h^*\sqrt{\pi}},
		& r \le 2\sqrt{\pi}h^*,
	\end{cases}
\end{equation*}
where the hydrodynamic interaction parameter is defined as
$h^* = a/\sqrt{\pi k_{\text{B}}T/\hat{H}}$ is the non-dimensional bead radius.

The total non-dimensional force acting on bead $\nu$ is given by
\begin{equation}
	\bm{F}_{\nu} = \bm{F}_{\nu}^{S} + \bm{F}_{\nu}^{SDK}.
\end{equation}
where the spring force $\bm{F}_{\nu}^{S}$ is modeled using a Hookean spring law and $\bm{F}_{\nu}^{SDK}$ is modeled using the Soddemann-D\"unweg-Kremer (SDK) potential\citep{Soddemann2001}. In dimensional form,
\begin{equation}
	\hat{\bm{F}}^{S} = \hat{H}\,\hat{\bm{Q}},
\end{equation}
where $\hat{\bm{Q}}$ is the connector vector between adjacent beads. In non-dimensional form, scaled using Hookean units, the spring force becomes
\begin{equation}
	\bm{F}^{S} = \bm{Q}.
\end{equation}

\begin{figure*}[!ht]
\begin{center}
\includegraphics[width=16cm,height=!]{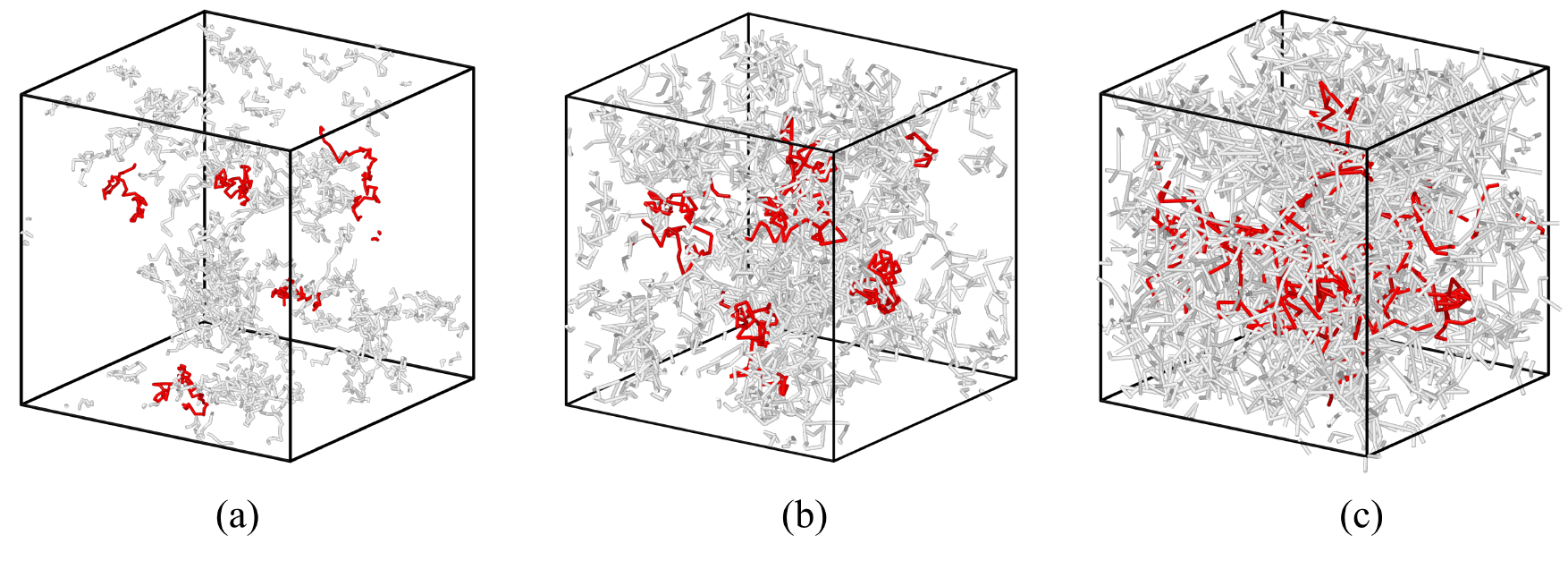}
\end{center}
\vskip-10pt
\caption{Simulation snapshots of flexible Hookean chains ($N_b=64$) in a $\theta$ solvent at concentrations (a) $c/c^{*}=0.1$, (b) $c/c^{*}=1$, and (c) $c/c^{*}=6$. Selected chains are highlighted in red for visual clarity, while all other chains are shown in gray.}
\label{fig:snapshots}
\vskip-10pt
\end{figure*}

The force ($\bm{F}_{\nu}^{SDK}$) due to excluded volume interactions between bead pairs is derived from the SDK potential:
{\small
\begin{align}\label{eq:SDK} 
\frac{\hat{U}_{\mu\nu}^{\text{SDK}}}{k_{\text{B}} T} =
\begin{cases}
4\!\left[\left(\dfrac{\sigma}{r_{\mu\nu}}\right)^{12}
- \left(\dfrac{\sigma}{r_{\mu\nu}}\right)^6 + \dfrac{1}{4}\right] - \epsilon,
& r_{\mu\nu} \le 2^{1/6} \sigma \\[20pt]
\dfrac{1}{2}\epsilon\!\left[\cos\!\left(\beta_1 \left(\dfrac{r_{\mu\nu}}{\sigma}\right)^2 + \beta_2\right) - 1\right],
& 2^{1/6} \sigma \le r_{\mu\nu} \\[-5pt] & r_{\mu\nu}\le r_{\mathrm{cutoff}} \\[15pt]
0, & r_{\mu\nu} \ge r_{\mathrm{cutoff}}
\end{cases}
\end{align}
}
Here, $\epsilon$ denotes the well depth of the SDK potential and controls the strength of interactions between bead pairs, while the non-dimensional bead diameter $\sigma$ is set to unity. The repulsive part of the interaction is described by a truncated Lennard-Jones (LJ) potential, whereas the attractive part is modeled using a cosine function. The constants $\beta_1$ and $\beta_2$ are determined from continuity and smoothness conditions, such that $\hat{U}_{\mu\nu}^{\text{SDK}} (r_{\mu\nu}) = 0$ at the cutoff distance $r_{\mu\nu} = r_{\mathrm{cutoff}}$ and $\hat{U}_{\mu\nu}^{\text{SDK}}(r_{\mu\nu}) = -\epsilon$ at $r_{\mu\nu} = 2^{1/6}\sigma$, corresponding to the minimum of the LJ potential. The cutoff radius is chosen as $r_{\mathrm{cutoff}} = 1.82\sigma$, following the discussion in recent work by \citet{Santra2019}. At $\epsilon = 0$, the SDK potential reduces to the purely repulsive Weeks-Chandler-Andersen (WCA) potential, corresponding to good solvent conditions in the athermal limit; this case will hereafter be referred to as athermal solutions. Within the context of the SDK potential, the solvent quality can be systematically varied by adjusting the value of $\epsilon$~\citep{Santra2019}. Here, attention is restricted to the limits of $\theta$ and athermal solvents. For $\theta$ solvent conditions, polymer chains are modeled by switching off excluded-volume interactions entirely, such that the SDK force vanishes, $\bm{F}_{\text{SDK}} = \bm{0}$.

\subsection{\label{sec:param} Simulation parameters}

Brownian dynamics simulations were performed using the HOOMD-blue simulation toolkit \citep{Anderson2020}. The decomposition of the diffusion tensor, required for incorporating hydrodynamic interactions, was carried out using the Positively Split Ewald (PSE) method. This approach was implemented as a plugin to HOOMD-blue by \citet{Fiore2017}. Although the original PSE algorithm was developed for colloidal suspensions, it has since been adapted for polymer solutions and shown to accurately capture hydrodynamic interactions in bead-spring polymer models \citep{Robe2024}, and this implementation is employed in the present study.

Polymer chains are modeled as Hookean bead-spring chains discretized into $N_b = 32$ to $96$ beads, depending on the system considered. Infinitely dilute solutions were simulated at a reduced concentration of $c/c^* = 10^{-6}$, ensuring the absence of intermolecular interactions. Finite-concentration simulations span a reduced concentration range of $c/c^* = 0.1 \, \text{--} \, 6$, covering both the dilute and semidilute unentangled regimes. The simulation box size was chosen such that the box length satisfied $L \approx 2.5\langle R_{\mathrm{e}}\rangle$, where $\langle R_{\mathrm{e}}\rangle$ is the equilibrium end-to-end distance of the polymer chain. Athermal solvent conditions were modeled using the SDK potential with $\epsilon = 0$, while $\theta$ solvent conditions were obtained by switching off excluded-volume interactions entirely. Fig.~\ref{fig:snapshots} shows representative simulation snapshots of semidilute solutions of flexible Hookean chains in a $\theta$ solvent at concentrations $c/c^{*}=0.1$, $1$, and $6$. For infinitely dilute solutions, a concentration of $c/c^{*}=10^{-6}$ was modeled by simulating a single polymer chain in a sufficiently large periodic simulation box, such that interactions with its periodic images were negligible.

To ensure that each trajectory reached equilibrium, the equilibration phase was monitored by tracking the time evolution of the radius of gyration until it attained a steady value. Simulations were performed using a non-dimensional time step $\Delta t = 10^{-3}$, and time step convergence was verified by confirming that further reduction in $\Delta t$ did not affect the measured dynamic properties. Dynamic properties were evaluated as functions of time during the production phase of each trajectory. Ensemble averages and statistical uncertainties were obtained by averaging over approximately $12\,000$ independent trajectories.

All simulations were performed on the Gadi supercomputer at the National Computational Infrastructure (NCI), Australia. Production simulations were carried out on compute nodes equipped with four NVIDIA Tesla V100-SXM2-32GB GPUs, dual 24-core Intel Xeon Platinum 8268 (Cascade Lake) 2.9~GHz processors, and 384~GB RAM per node. The Brownian dynamics simulations were performed in an embarrassingly parallel manner, with statistically independent trajectories distributed across the four GPUs, each GPU executing one trajectory independently without requiring communication or data transfer between GPUs. The most computationally demanding simulations corresponded to $c/c^*=6$ and simulation box length $L = 42.43 l_H$, with $N_b=96$, containing $N_c=32$ chains ($N=N_cN_b=30,720$ beads) within a simulation box. Each independent run, or independent simulation box was evolved for $12,000 \lambda_H$, and $600$ statistically independent runs were used to obtain the reported averages. A single independent run required approximately 40 minutes on one NVIDIA Tesla V100 GPU. The data for a single $G(t)$ curve for $N_b=96$ in Fig.~\ref{fig:dilute}(a) required around $100$ hours wall-clock time when run in $4$ NVIDIA Tesla V100 GPUs.

\subsection{\label{sec:properties} Estimating linear viscoelastic properties}

In this section we will discuss the procedure to calculate various linear viscoelastic properties from the results of Brownian dynamics simulations.

\subsubsection{Stress Tensor}
The dynamic properties such as relaxation modulus, zero-shear rate viscosity and dynamic moduli investigated in this work can be defined in terms of the components of the stress-tensor \citep{Bird1987} for the polymer solution. The instantaneous non-dimensional contribution to the polymer stress tensor at a given time is given by the Kramers--Kirkwood expression,
\begin{equation}\label{Kramers}
\bm{\tau}_{\text{p}} = \frac{1}{N_{c}} \left\langle \sum_{\xi=1}^{N_{c}} \sum_{\nu=1}^{N_b} \left(\bm{r}_\nu^{(\xi)}-\bm{r}_c^{(\xi)} \right) \bm{F}_{\xi\nu} \right\rangle
\end{equation}
where the stress is non-dimensionalised by $N_{c} \, (k_B T/V)$ and $\bm{r}_c$ is the center-of-mass position vector. The force on each bead $\mu$ in a chain $\xi$ is given by,
\begin{equation}\label{totalforce}
\bm{F}_{\xi\nu} = \sum_{\beta=1}^{N_{c}} \sum_{\substack{\mu=1 \\ \mu\ne\nu}}^{N_b} \bm{F}_{\xi\nu,\beta\mu}^{\text{SDK}} + \sum_{\substack{\mu=1 \\ \mu\ne\nu}}^{N_b} \bm{F}_{\xi\nu,\xi\mu}^{\text{S}}
\end{equation}
The instantaneous tensor $\bm{S}$, which is the term inside the brackets in the non-dimensional polymer stress tensor in Eq.~\ref{Kramers}, represents the stress contribution from all polymer chains in a single simulation box, is defined by
\begin{equation}\label{S}
\bm{S} = \sum_{\xi=1}^{N_c} \sum_{\nu=1}^{N_b} \left( \bm{r}_\nu^{(\xi)}-\bm{r}_c^{(\xi)} \right) \bm{F}_{\xi\nu}
\end{equation}
which can be simplified to,
\begin{align}
\label{Kramersv2}
\bm{S} &=  
\frac{1}{2} \sum_{\nu=1}^{N}\sum_{\substack{\mu=1 \\ \mu\ne\nu}}^{N} 
\bm{r}_{\nu\mu} \bm{F}_{\nu\mu}^{\text{SDK}} 
+ \sum_{\xi=1}^{{N}_{c}} \sum_{i=1}^{N_b-1} 
\bm{Q}_{i}^{(\xi)} \bm{F}^{{\text{S}}}\!\left( \bm{Q}_{i}^{(\xi)} \right).
\end{align}

Once the stress tensor is computed, we can easily estimate various dynamic properties and material functions for the polymer solution. Here, the focus is on calculating linear viscoelastic properties in terms of the relaxation modulus, the zero shear rate viscosity, and dynamic moduli.

\subsubsection{Relaxation Modulus}
The relaxation modulus \(G(t)\) is obtained from equilibrium simulations  using the Green-Kubo relation that relates \(G(t)\) to the autocorrelation function of the stress tensor \citep{Wittmer2015}. At equilibrium, the stress tensor is isotropic, hence the relaxation modulus is given by the expression,
\begin{equation}
G(t) = \frac{1}{3} \left( G_{xy}(t) + G_{xz}(t) + G_{yz}(t) \right)
\end{equation}
where the components \(G_{ij}(t)\) (which are equal to each other at equilibrium) are given by the expression,
\begin{equation}
G_{ij}(t) = \frac{1}{N_c} \left\langle S_{ij}(0) \, S_{ij}(t) \right\rangle
\end{equation}
The relaxation modulus can be easily computed using the above equation.

The relaxation modulus obtained from the simulations is fitted with a sum of exponentials \citep{Pan2021, Varakhedkar2025},
\begin{equation}
G(t) = \sum_{i=1}^{n} a_i \exp(- b_i t)
\label{eq:sum_exp_fit}
\end{equation}
where $a_i$ and $b_i$ are the fitting parameters and $n$ is the number of exponentials required to fit the curve. All the relaxation modulus curves evaluated in this paper are fitted using $5$ to $9$ exponentials. The errorbars of the fitted $G(t)$ is obtained from the envelope of curves reconstructed using the upper and lower bounds of the coefficients. The relaxation modulus is fitted using a nonlinear least-squares procedure with initial guesses $a_i=1$ and $b_i=100$. To assess the sensitivity of the fit to the initial parameter estimates, the fitting procedure was repeated using initial guesses in the ranges $a_i=0.1-10$ and $b_i=10-1000$. The resulting fitted parameters exhibited only minor variations, leading to changes of less than $1\%$ in the fitted response, indicating that the fit is insensitive to the choice of initial guess parameters. All the subsequent calculations involving the relaxation modulus are carried out using similar fits.

\subsubsection{Zero Shear Rate Viscosity}
While the study of shear viscosity at moderately high shear rates \(\dot{\gamma}\) is important in non-linear rheology, the study of linear viscoelasticity primarily focuses on the polymeric component of the zero-shear rate viscosity, defined as: $ \eta_{p,0} = \lim_{\dot{\gamma} \to 0} \eta_p$. Here \(\eta_{p,0} \) is calculated from equilibrium simulations by integrating the relaxation modulus $G(t)$ \citep{Fixman1981, Lee2009},
\begin{equation}
\eta_{p,0} = \int_0^\infty G(t) \, dt
\end{equation}
Substituting Eq.~\ref{eq:sum_exp_fit} into the definition of the zero-shear viscosity above  yields, $\eta_{p,0}=\sum_i a_i b_i$, where $a_i$ and $b_i$ are the fitted coefficients and relaxation times, respectively.

\subsubsection{Dynamic Moduli}
The elastic and viscous response of a viscoelastic fluid is generally characterised by the storage \(G'\) and the loss \(G''\) moduli, together referred to as the dynamic moduli. These properties are typically obtained from oscillatory shear flow experiments. In the current simulations, in the limit of very small strain amplitude, \(G'\) and \(G''\) are determined from a Fourier transformation of the relaxation modulus $G(t)$ \citep{Wittmer2015}, 
\begin{equation}
G'(\omega) = \int_{0}^{\infty} G(t) \sin(\omega t) \, d(\omega t)
\end{equation}
\begin{equation}
G''(\omega) = \int_{0}^{\infty} G(t) \cos(\omega t) \, d(\omega t)
\end{equation}
Alternatively, dynamic moduli can also be obtained by performing the Fourier transform on the analytical representation of $G(t)$ provided by the sum-of-exponentials fit in Eq.~\ref{eq:sum_exp_fit}.

\section{Results and discussion}

\begin{figure}[t]
\begin{center}
\begin{tabular}{c}
        \includegraphics[width=8.35cm,height=!]{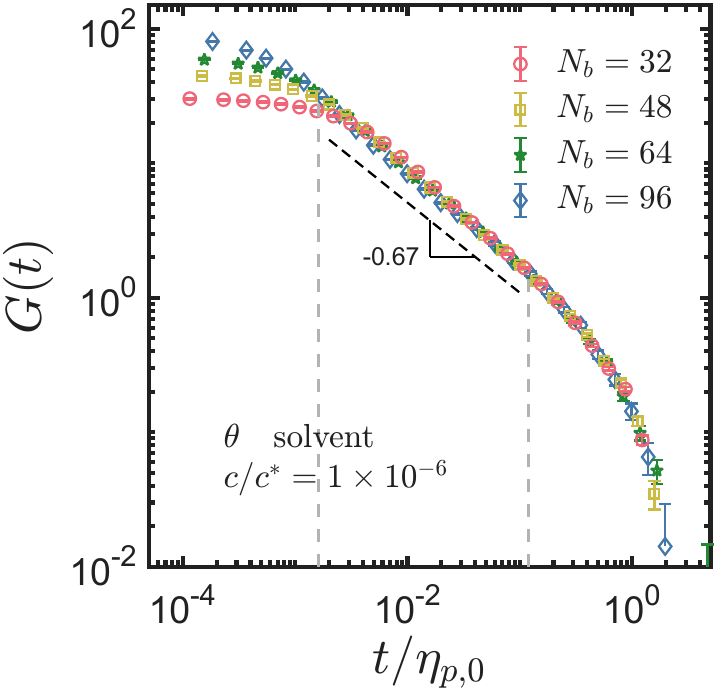} \\
          \large{(a)}\\[5pt] 
        \includegraphics[width=8.35cm,height=!]{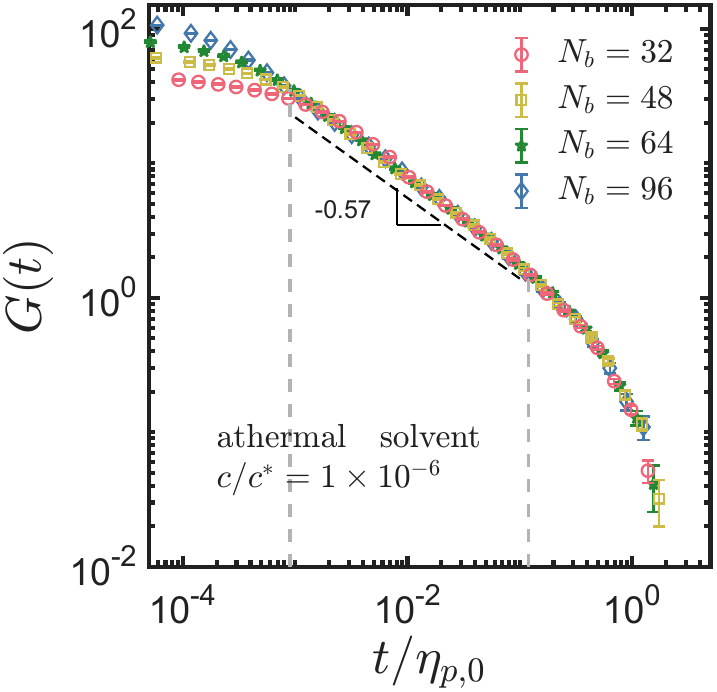} \\
             \large{(b)} 
\end{tabular}
\end{center}
\vskip-15pt
	\caption{Non-dimensional relaxation modulus $G(t)$ as a function of non-dimensional time scaled using zero-shear viscosity $\eta_{p,0}$. Flexible polymer chains in infinitely dilute solution are simulated for varying discretization $N_b = 32, 48, 64,$ and $96$ for (a) $\theta$ solvent and (b) athermal solvent. Here the slopes represent the Zimm power law exponents for $\theta$ and athermal solvent. The vertical dashed grey lines indicate the intermediate-time window over which the power-law exponent $\alpha$ was obtained by fitting $G(t)\sim t^{-\alpha}$ for $N_b=96$.}
	\label{fig:dilute}
	\vskip-20pt
\end{figure}

\subsection{\label{sec:dilute} Dilute solutions}

\begin{figure*}[t]
\begin{center}
\begin{tabular}{cc}
        \includegraphics[width=8.35cm,height=!]{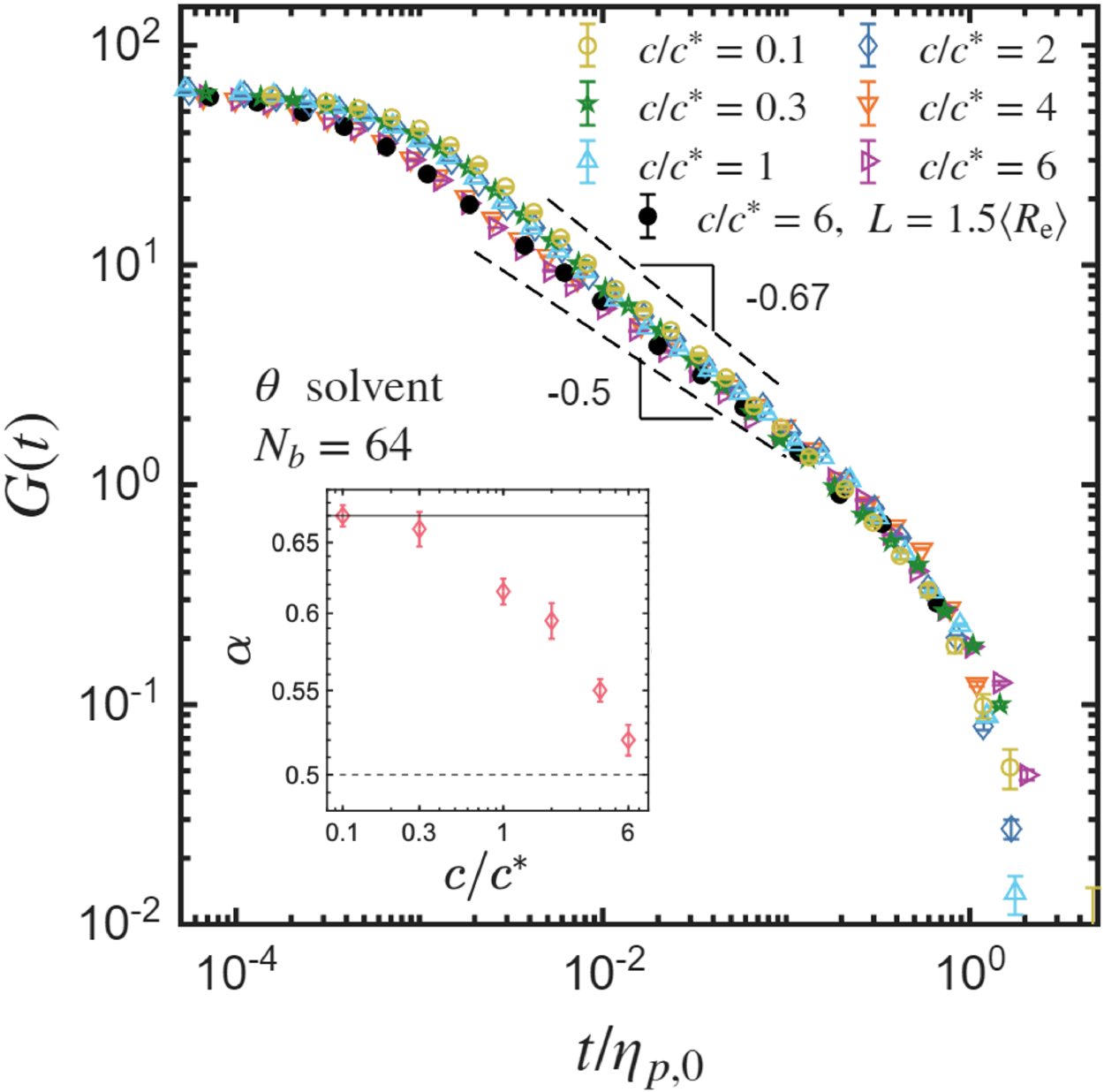} &
        \includegraphics[width=8.35cm,height=!]{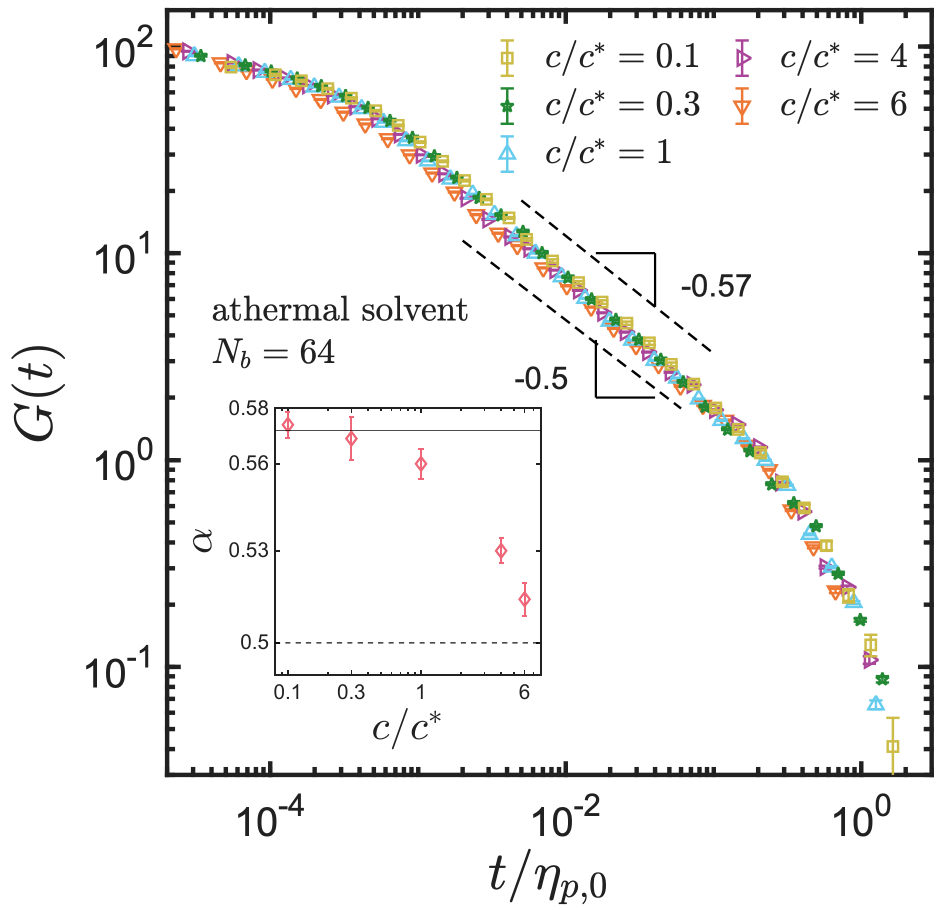} \\
             \large{(a)}&   \large{(b)} 
\end{tabular}
\end{center}
\vskip-15pt
	\caption{Non-dimensional relaxation modulus $G(t)$ as a function of non-dimensional time scaled using the zero-shear viscosity $\eta_{p,0}$. Flexible polymer chains with $N_b = 64$ are simulated at varying concentrations $c/c^{*} = 0.1$ to $6$ for (a) $\theta$ solvent and (b) athermal solvent. The slopes in the main panels indicate the Rouse and Zimm power-law exponents. Insets show the variation of the intermediate scaling exponent $\alpha$ with concentration $c/c^{*}$ for (a) $\theta$ solvent and (b) athermal solvent. All simulations shown were performed using a simulation box of size $L \approx 2.5\langle R_{\mathrm{e}}\rangle$. The black filled symbols correspond to $c/c^{*}=6$ obtained using a smaller simulation box, $L \approx 1.5\langle R_{\mathrm{e}}\rangle$. The solid and dashed lines correspond to the scaling exponents predicted by the Zimm and Rouse theories, respectively.}
	\label{fig:semidilute_Gt}
\vskip-10pt
\end{figure*}

The linear viscoelastic response of flexible polymer chains in the infinitely dilute limit provides a baseline for validating the simulation framework. Fig.~\ref{fig:dilute} shows the results of Brownian dynamics simulations for stress relaxation modulus $G(t)$ for chains with $N_b = 32$ to $96$. Results are shown for $\theta$ solvent conditions (Fig.~\ref{fig:dilute}(a)) and for athermal solvent conditions (Fig.~\ref{fig:dilute}(b)). In both cases, data corresponding to different chain lengths collapse onto a single master curve once time is scaled using zero-shear viscosity $\eta_{p,0}$ obtained by integrating the fitted relaxation modulus with time, indicating that the relaxation modulus is independent of $N_b$ once appropriate normalization is employed. The corresponding zero-shear viscosities used to normalize the time axis are reported in Table~\ref{tab:dilute_viscosity}. In addition to the simulation values obtained for $\theta$ and athermal solvent conditions, the table also lists the corresponding Zimm-model predictions and the ratio $R_0=\eta_{p,0}^{\mathrm{sim}}/\eta_{p,0}^{\mathrm{Zimm}}$. Since the original Zimm theory is formulated for $\theta$ solvents, the values of $\eta_{p,0}^{\mathrm{Zimm}}$ reported in Table~\ref{tab:dilute_viscosity} correspond to the $\theta$-solvent case and are used as the reference values in evaluating $R_0$ for both solvent conditions. The ratio $R_0$ is discussed in more detail susequently.

\begin{table}[t]
\caption{Comparison of zero-shear viscosities in the dilute regime. The ratio $R_0=\eta_{p,0}^{\mathrm{sim}}/\eta_{p,0}^{\mathrm{Zimm}}$ is reported for both $\theta$ and athermal solvent conditions.}
\label{tab:dilute_viscosity}
\centering
\renewcommand{\arraystretch}{1.3}
\setlength{\tabcolsep}{3pt}
\begin{tabular}{ccccccc}
\hline
\multirow{2}{*}{$N_b$} &
\multirow{2}{*}{$\eta_{p,0}^{\mathrm{Zimm}}$} &
\multicolumn{2}{c}{$\eta_{p,0}^{\mathrm{sim}}$} &
\multicolumn{2}{c}{$R_0$} \\
\cline{3-4}\cline{5-6}
&
& $\theta$
& Athermal
& $\theta$
& Athermal \\
\hline
32 &
210.6 &
$173.7 \pm 6$ &
$436.6 \pm 8$ &
0.825 &
2.073 \\
48 &
402.0 &
$338.9 \pm 9$ &
$871.2 \pm 14$ &
0.843 &
2.167 \\
64 &
633.5 &
$515.1 \pm 11$ &
$1559.2 \pm 18$ &
0.813 &
2.461 \\
80 &
899.6 &
$766.1 \pm 15$ &
$2084.9 \pm 21$ &
0.852 &
2.317 \\
96 &
1196.8 &
$980.5 \pm 18$ &
$3051.2 \pm 25$ &
0.819 &
2.549 \\
\hline
\end{tabular}
\vskip-10pt
\end{table}

In the intermediate-time regime, the relaxation modulus exhibits a power-law decay of the form $G(t)\sim t^{-\alpha}$, where $\alpha$ denotes the corresponding power-law exponent. The exponent $\alpha$ is determined by fitting the intermediate-time power-law regime of the relaxation modulus, lying between the short-time bead-spring plateau and the long-time exponential decay. Representative fitting windows for $N_b=96$ under both $\theta$ and athermal solvent conditions are indicated by the vertical dashed lines in Fig.~\ref{fig:dilute}. Within the Zimm framework, the scaling of $G(t)$ is governed by the Flory exponent $\nu$, with $G(t) \sim t^{-1/(3\nu)}$ \citep{Doi1988,Rubinstein2003}. For $\theta$ solvents, where $\nu = 1/2$, this yields $G(t) \sim t^{-2/3}$, while for athermal solvents, where $\nu \approx 0.588$, the corresponding prediction is $G(t) \sim t^{-0.57}$. The observed scaling behavior in both solvent conditions is consistent with these theoretical predictions, reflecting the role of excluded-volume interactions in modifying the relaxation spectrum in good solvents. It is worth noting that power law scaling is observed over a longer period at intermediate times as $N_b$ increases. Scaling predictions are valid in the limit of long chains. These results establish that the simulations reproduce the expected dilute-solution viscoelastic behavior for flexible polymer chains, providing a consistent reference for the analysis of semidilute solutions presented in subsequent sections.

\subsection{\label{sec:semidilute} Semidilute solutions}

In this section, the linear viscoelastic response of flexible polymers in the semidilute unentangled regime is examined, where interchain interactions and hydrodynamic screening become important. Fig.~\ref{fig:semidilute_Gt} shows the stress relaxation modulus $G(t)$ for chains of fixed length $N_b = 64$ at concentrations ranging from $c/c^* = 0.1$ to $6$, for both $\theta$ solvent conditions (Fig.~\ref{fig:semidilute_Gt}(a)) and athermal solvent conditions (Fig.~\ref{fig:semidilute_Gt}(b)). The figure also includes results for $c/c^{*}=6$ obtained using two different simulation box sizes, corresponding to different numbers of polymer chains. The data obtained using the smaller simulation box ($L \approx 1.5\langle R_{\mathrm{e}}\rangle$) are indistinguishable from those obtained using the standard box size ($L \approx 2.5\langle R_{\mathrm{e}}\rangle$), indicating that the viscoelastic response is insensitive to the number of chains used in the simulation for the system sizes considered. The corresponding zero-shear rate viscosities used to normalize the time axis are reported in Table~\ref{tab:semidilute_viscosity}. The table lists the simulation values of $\eta_{p,0}$ for both $\theta$ and athermal solvent conditions together with the ratio $R_0=\eta_{p,0}^{\mathrm{sim}}/\eta_{p,0}^{\mathrm{Zimm}}$, where $\eta_{p,0}^{\mathrm{Zimm}}$ is the Zimm prediction for a chain with $N_b=64$.

\begin{table}[t]
\caption{Comparison of zero-shear viscosities in the semidilute regime for chains with $N_b=64$. The ratio $R_0=\eta_{p,0}^{\mathrm{sim}}/\eta_{p,0}^{\mathrm{Zimm}}$ is reported for both solvent conditions.}
\label{tab:semidilute_viscosity}
\centering
\renewcommand{\arraystretch}{1.3}
\setlength{\tabcolsep}{3pt}
\begin{tabular}{ccccc}
\hline
\multirow{2}{*}{$c/c^*$} &
\multicolumn{2}{c}{$\eta_{p,0}^{\mathrm{sim}}$} &
\multicolumn{2}{c}{$R_0$} \\
\cline{2-3}\cline{4-5}
&
$\theta$ &
Athermal &
$\theta$ &
Athermal \\
\hline
0.1 &
$526.1 \pm 8$ &
$1476.8 \pm 15$ &
0.830 &
2.331 \\
0.3 &
$583.9 \pm 10$ &
$1559.2 \pm 17$ &
0.922 &
2.461 \\
1 &
$685.0 \pm 12$ &
$1635.0 \pm 18$ &
1.081 &
2.581 \\
2 &
$818.9 \pm 14$ &
--- &
1.292 &
--- \\
4 &
$1176.2 \pm 19$ &
$1811.0 \pm 21$ &
1.857 &
2.859 \\
6 &
$1213.6 \pm 21$ &
$2173.1 \pm 24$ &
1.916 &
3.430 \\
\hline
\end{tabular}
\vskip-10pt
\end{table}

At low concentrations, the intermediate-time behavior closely resembles that observed in the dilute limit, reflecting Zimm-like dynamics. As the concentration increases, a systematic crossover in the intermediate-time scaling is observed, with the effective power-law exponent decreasing from its dilute-solution value toward Rouse-like behavior characterized by $t^{-0.5}$. In the $\theta$ solvent case, the scaling transitions from $t^{-0.67}$ to $t^{-0.5}$, while in the athermal solvent case the corresponding crossover is from $t^{-0.57}$ to $t^{-0.5}$. The concentration dependence of the intermediate-time exponent is further quantified in the inset plots (Fig.~\ref{fig:semidilute_Gt}) of $\alpha$ as a function of $c/c^*$ for both solvent conditions. In both cases, $\alpha$ decreases monotonically with increasing concentration, evolving smoothly from Zimm to Rouse limit of $\alpha = 0.5$. Blob theory identifies the correlation blob as the characteristic length scale below which hydrodynamic and excluded-volume interactions remain unscreened and above which they are screened, thereby predicting the asymptotic Zimm and Rouse scaling regimes. As the polymer concentration increases, the blob size decreases, so that progressively smaller segments of a polymer chain experience unscreened hydrodynamic and excluded-volume interactions, while interactions over larger length scales become increasingly screened. In this sense, the concentration-dependent evolution of the intermediate power-law exponent $\alpha$ observed in the present simulations is consistent with the progressive hydrodynamic screening mechanism underlying blob theory. While blob theory does not predict how the intermediate exponent evolves during the crossover, the present Brownian dynamics simulations quantitatively resolve its continuous evolution with increasing concentration.

\begin{figure}[!t]
\begin{center}
\begin{tabular}{c}
        \includegraphics[width=8.35cm,height=!]{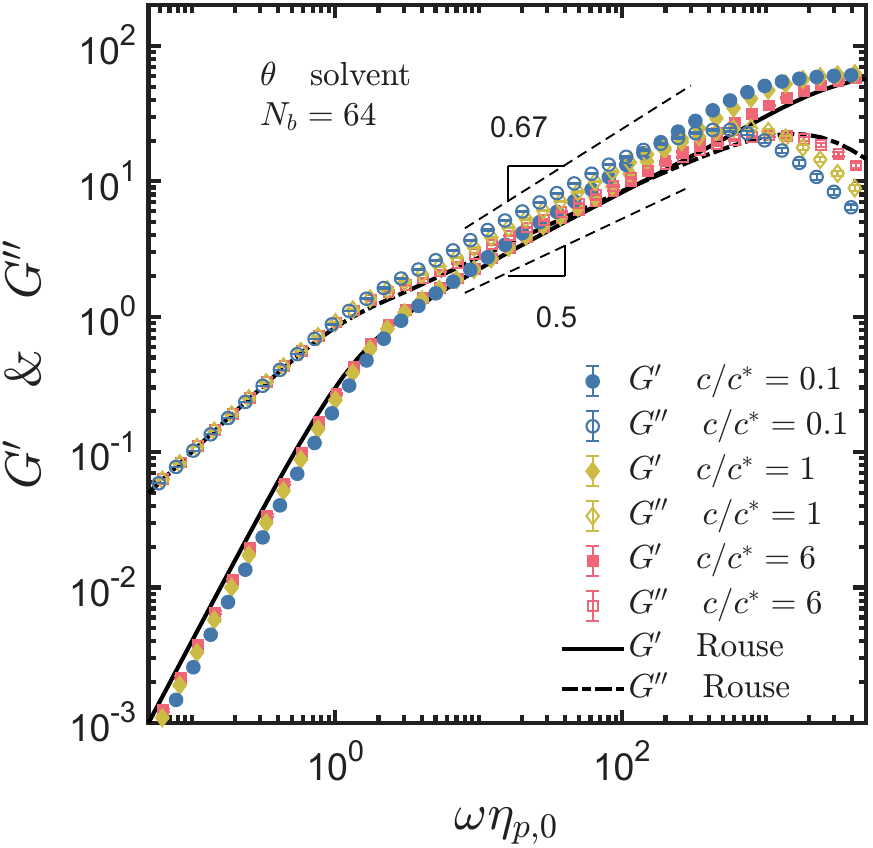} \\
          \large{(a)}\\[5pt] 
        \includegraphics[width=8.35cm,height=!]{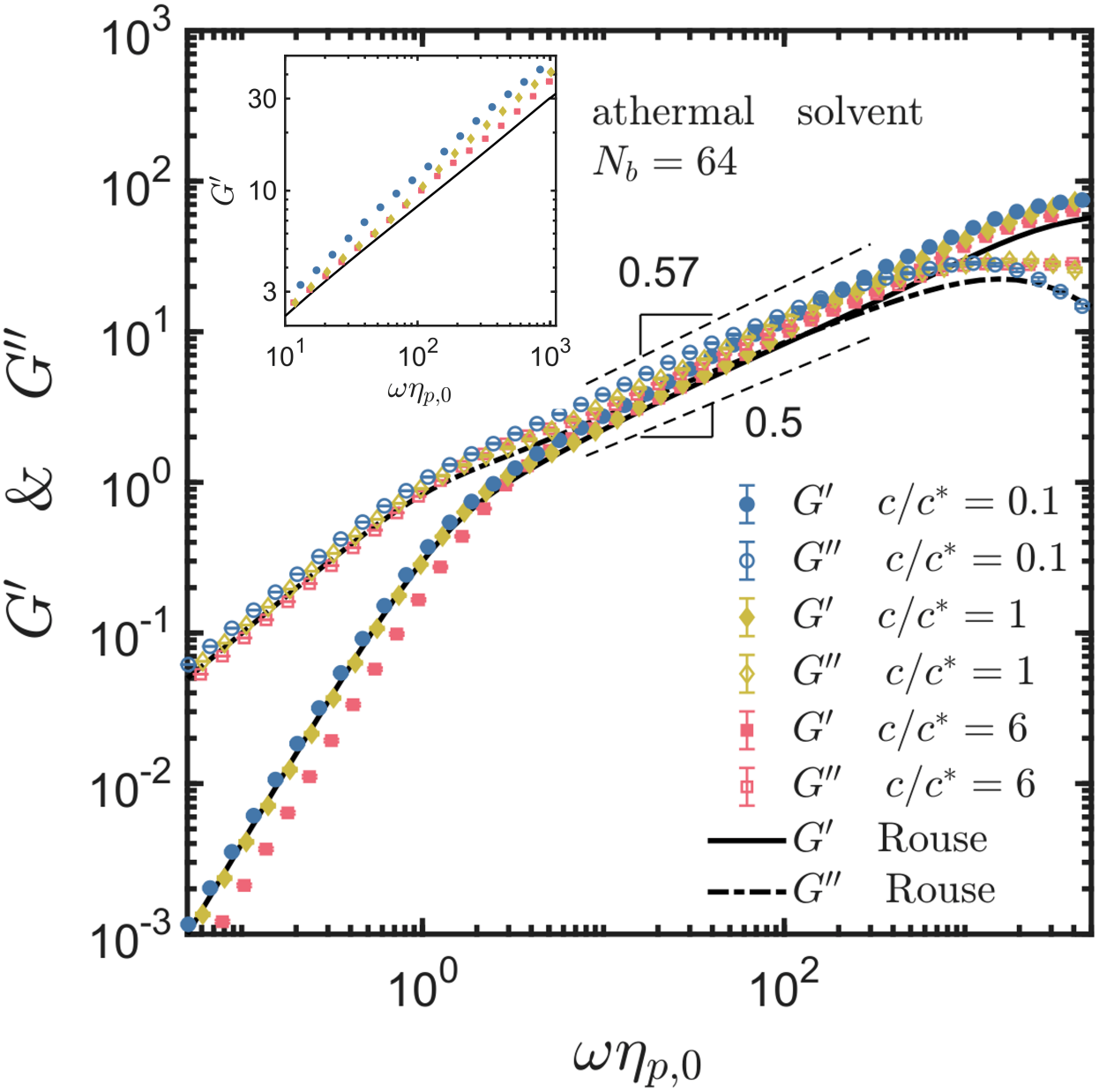} \\
             \large{(b)} 
\end{tabular}
\end{center}
\vskip-15pt
	\caption{Non-dimensional dynamic moduli, $G'$ (filled) and $G''$ (hollow) as a function of frequency scaled with zero-shear viscosity $\eta_{p,0}$. Flexible polymer chains with $N_b = 64$ are simulated at varying concentrations $c/c^{*} = 0.1, 1$ and $6$ for (a) $\theta$ solvent and (b) athermal solvent. The black lines are the dynamic moduli of the Rouse analytical model for $N_b=64$ \citep{Bird1987}. Here the slopes represent the Rouse and Zimm power law exponents for $\theta$ and athermal solvent. The inset in (b) shows $G'(\omega)$ over a selected frequency range to highlight the change in slope with increasing concentration.}
	\label{fig:semidilute_Gw}
	\vskip-10pt
\end{figure}

Hydrodynamic screening in semidilute polymer solutions is also associated with a characteristic timescale. \citet{Ahlrichs2001} demonstrated that hydrodynamic interactions remain unscreened at short times and become screened only after a crossover time corresponding to the Zimm relaxation time of a correlation blob. For a $\theta$ solvent, the corresponding scaling estimate is $t_c/\tau_Z \sim (c/c^*)^{-3}$, where $\tau_Z$ is the longest Zimm relaxation time of the chain. At the highest concentration considered here ($c/c^*=6$), this estimate gives $t_c/\tau_Z \sim 5\times10^{-3}$, corresponding to a characteristic reduced frequency $\omega_c\tau_Z \sim 200$. For frequencies below $\omega_c$, hydrodynamic interactions are screened and the dynamics are therefore expected to exhibit Rouse-like behavior, whereas for frequencies above $\omega_c$, hydrodynamic interactions remain unscreened and Zimm-like dynamics should, in principle, be observed. On the other hand, for the longest chain considered in the present work ($N_b=96$), the relaxation time of the fastest internal mode is approximately $2\times10^{-3}\tau_Z$, corresponding to a limiting reduced frequency $\omega_f\tau_Z \approx 500$, beyond which the finite chain no longer exhibits the intermediate-frequency scaling regime. Since $\omega_c$ and $\omega_f$ are of the same order, the frequency window over which unscreened hydrodynamic interactions could be observed is expected to be very narrow for the present chain discretizations. Consequently, the crossover is manifested primarily through the smooth evolution of the effective intermediate-time scaling exponent rather than as a distinctly resolved transition between the asymptotic Zimm and Rouse scaling regimes. Longer chains would separate these characteristic frequencies and enable the crossover in $G'(\omega)$ and $G''(\omega)$ to be resolved more clearly.

The concentration-induced crossover observed in the time domain is also evident in the frequency-dependent dynamic moduli. Fig.~\ref{fig:semidilute_Gw} shows the storage and loss moduli, $G'(\omega)$ and $G''(\omega)$, for $c/c^*=0.1$, $1$, and $6$, together with the dynamic moduli obtained from the analytical Rouse model for a chain of length $N_b=64$ \citep{Bird1987}. At low concentrations, both $G'$ and $G''$ exhibit intermediate-frequency scaling consistent with Zimm-like dynamics, while at higher concentrations the moduli progressively approach the Rouse behavior. In particular, at $c/c^*=6$, the simulated data closely follow the Rouse curves over a broad frequency range, demonstrating that, with an increase in concentration, hydrodynamic interactions are sufficiently screened.

\begin{figure*}[!ht]
	\begin{center}
		\begin{tabular}{cc}
			\includegraphics[width=8.35cm,height=!]{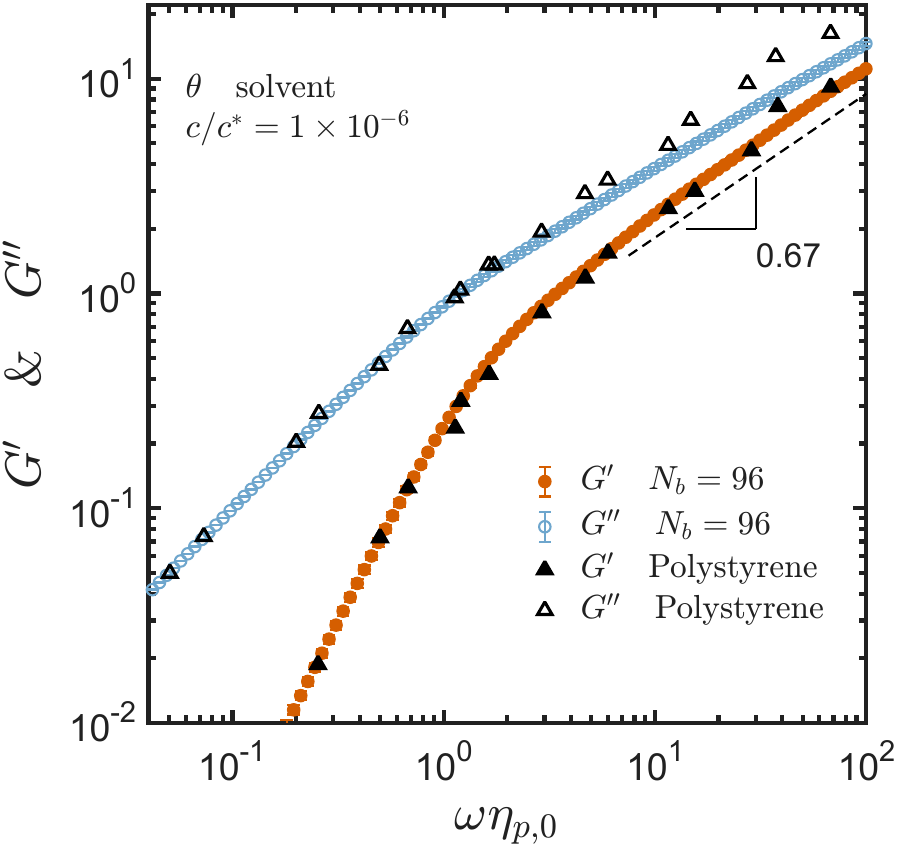} &
			\includegraphics[width=8.35cm,height=!]{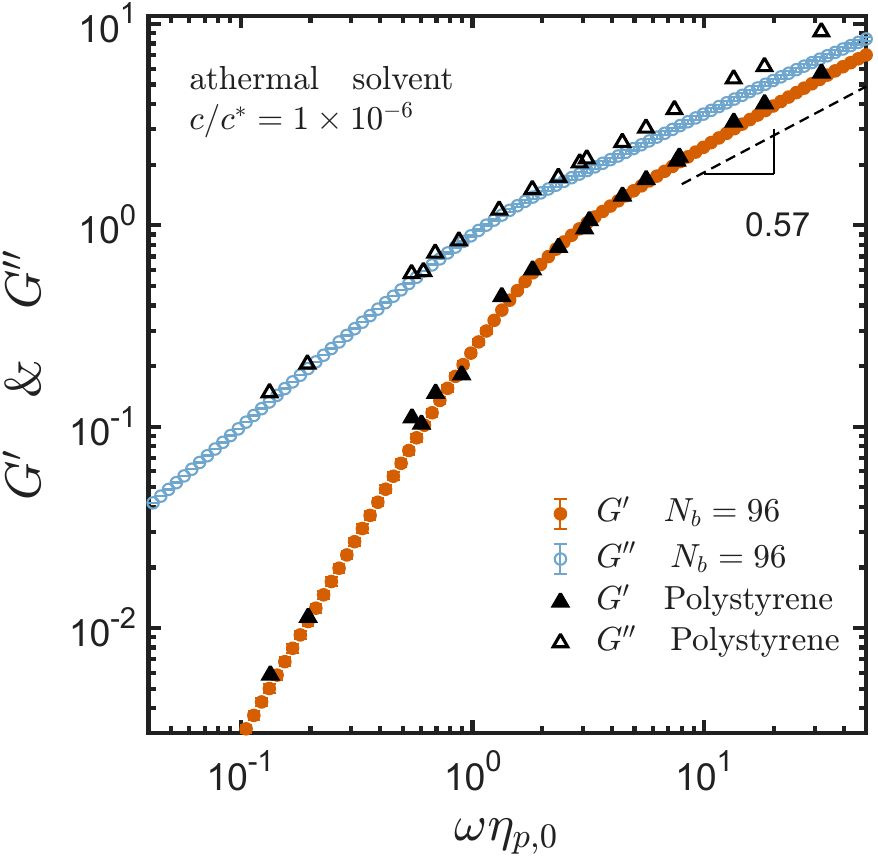} \\
			\large{(a)}&   \large{(b)} 
		\end{tabular}
	\end{center}\vskip-15pt
	\caption{Comparison of dynamic moduli with the experimental data for polystyrene in an infinitely dilute solution in (a) $\theta$ solvent and (b) athermal solvent. The non-dimensional dynamic moduli, $G'(\omega)$ (filled) and $G''(\omega)$ (hollow) are plotted as a function of frequency scaled with zero-shear viscosity $\eta_{p,0}$. Experimental data for polystyrene is taken from \citet{Johnson1970}. Simulation results correspond to chains of discretization $N_b=96$. Here the slopes represent the Zimm power law exponents for $\theta$ and athermal solvents.}
	\label{fig:exp_dilute}
\end{figure*}

\renewcommand{\arraystretch}{1.2}
\setlength{\tabcolsep}{6pt}
\begin{table*}[!t]
	\centering
	\caption{{Experimental polymer systems used for comparison with the present Brownian dynamics simulations. The parameters related to viscosity used for frequency normalization is listed for each experimental system.}}
	\vskip10pt
	\label{tab:exp_systems}
	\begin{tabular}{l l c c c c}
		\hline
		Polymer &
		Solvent &
		Solvent &
		$M_w$ &
		$c/c^*$ &
		Viscosity \\
		&
		&
		quality &
		(g/mol) &
		&
		(Pa\,s) \\
		\hline
		Polystyrene\citep{Johnson1970} &
		Decalin &
		$\theta$ &
		$8.6\times10^5$ &
		0 &
		$\eta_s[\eta]=0.224$ \\
		Polystyrene\citep{Johnson1970} &
		Di-2-ethylhexyl phthalate &
		$\theta$ &
		$8.6\times10^5$ &
		0 &
		$\eta_s[\eta]=5.15$ \\
		Polystyrene\citep{Johnson1970} &
		$\alpha$-chloronaphthalene &
		Athermal &
		$8.6\times10^5$ &
		0 &
		$\eta_s[\eta]=0.626$ \\
		Polystyrene\citep{Johnson1970} &
		Aroclor 1232 &
		Athermal &
		$8.6\times10^5$ &
		0 &
		$\eta_s[\eta]=2.60$ \\
		Poly(AN-co-IA)\citep{Zhu2012} &
		[BMIM]Cl &
		$\theta$ &
		$8.6\times10^4$ &
		0.78 &
		$\eta_{p,0}=0.28$ \\
		Poly(AN-co-IA)\citep{Zhu2012} &
		[BMIM]Cl &
		$\theta$ &
		$8.6\times10^4$ &
		2.22 &
		$\eta_{p,0}=0.77$ \\
		Poly(AN-co-IA)\citep{Zhu2012} &
		[BMIM]Cl &
		$\theta$ &
		$8.6\times10^4$ &
		5.56 &
		$\eta_{p,0}=5.19$ \\
		\hline
	\end{tabular}
\end{table*}

While the storage modulus $G'(\omega)$ follows the expected scaling behavior across all concentrations and smoothly approaches the Rouse prediction at high $c/c^*$, the loss modulus $G''(\omega)$ exhibits a departure from simple power-law scaling at high frequencies, reaching a maximum and subsequently decaying to zero. This feature is not a physical signature of additional relaxation mechanisms, but rather a numerical artefact arising from the discrete nature of the polymer model. For a finite chain represented by a discrete set of beads and springs, the relaxation spectrum contains a finite number of internal modes, with the highest-frequency modes associated with local bead-scale motion \citep{Rubinstein2003}. At frequencies comparable to or larger than the relaxation rate of the fastest internal mode, the storage modulus $G'(\omega)$ plateaus, whereas the loss modulus $G''(\omega)$ saturates and subsequently decays to zero. The same effect is present in the analytical Rouse model when evaluated for a finite number of modes, and therefore appears both in simulations and in the finite-$N_b$ Rouse prediction displayed in Fig.~\ref{fig:semidilute_Gw}. The consistency between simulations and the finite-chain Rouse model confirms that the observed deviations in $G''(\omega)$ originate from finite discretization rather than from limitations of the present model.

\subsection{\label{sec:exp_dil} Comparison with experiments: Dilute solutions}

Experimental measurements of dynamic moduli for flexible polymer chains in the limit of infinite dilution have been reported by \citet{Johnson1970} for high–molecular-weight polystyrene dissolved in $\theta$ solvents (decalin and di-2-ethylhexyl phthalate) and athermal solvents ($\alpha$-chloronaphthalene and Aroclor~1232) by extrapolating concentration-dependent data to the limit of zero polymer concentration. Fig.~\ref{fig:exp_dilute}~(a) displays their measurements of storage and loss moduli, $G'(\omega)$ and $G''(\omega)$, under $\theta$ conditions, while Fig.~\ref{fig:exp_dilute}~(b) displays their data for athermal solvent conditions~\citep{Johnson1970}. For the infinitely dilute polystyrene data of \citet{Johnson1970}, the reduced frequency is expressed in the original form, $\omega\eta_s[\eta]M/(N_Ak_BTS_p)$, where $\eta_s$ is the solvent viscosity, $[\eta]$ is the intrinsic viscosity, and $S_p=\sum_{p=1}^{N}\tau_p/\tau_1$ is the dimensionless sum of the normalized mode relaxation times. The values reported by \citet{Johnson1970} are $S_p=2.369$ for $\theta$ solvent conditions and $S_p=2.166$ for athermal solvent conditions. Since the polymer contribution to the zero-shear viscosity for dilute polymer solutions is given by $\eta_{p,0}=c\eta_s[\eta]$, corresponding to the leading-order term in the Taylor series expansion of the solution viscosity in powers of concentration, this normalization is consistent with the reduced frequency $\omega\eta_{p,0}M/(cN_Ak_BT\,S_p)$ adopted in the present work.

Table~\ref{tab:exp_systems} summarizes the molecular weights and the parameters related to viscosity used for frequency normalization for the experimental systems considered in the present work. For the infinitely dilute experiments of \citet{Johnson1970}, the viscosity parameter is reported as $\eta_s[\eta]$, following the original work. For the semidilute experiments of \citet{Zhu2012}, the experimentally measured polymer contribution to the zero-shear viscosity, $\eta_{p,0}$, is reported.

\begin{figure}[!ht]
	\begin{center}
\begin{tabular}{c}
        \includegraphics[width=8.35cm,height=!]{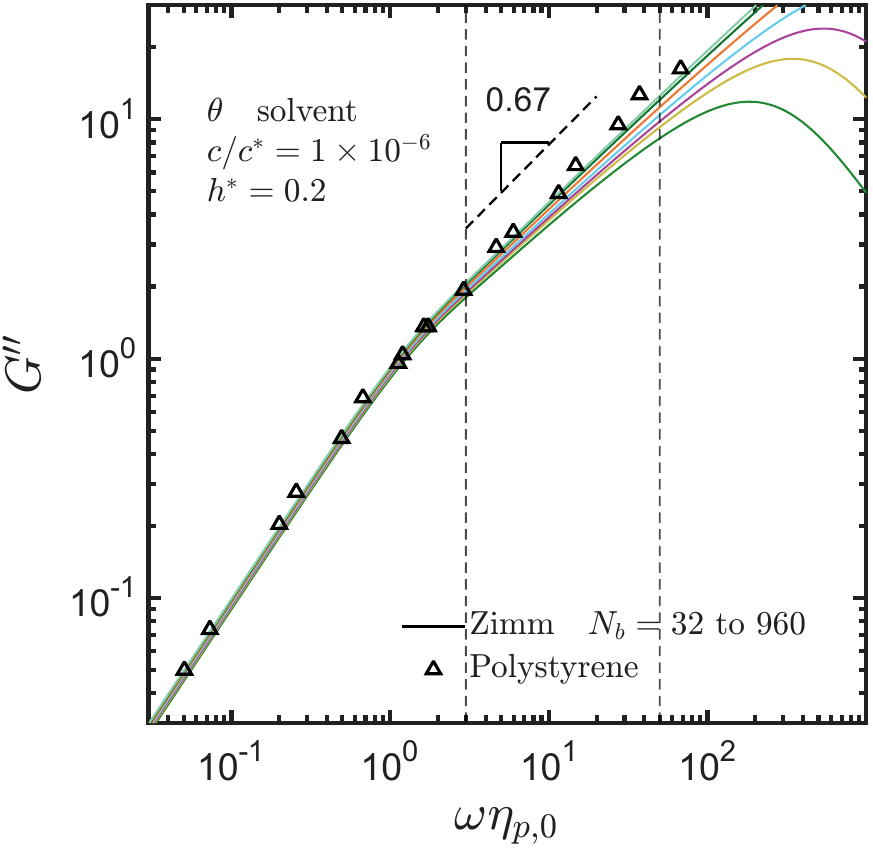} \\
          \large{(a)}\\[5pt] 
        \includegraphics[width=8.35cm,height=!]{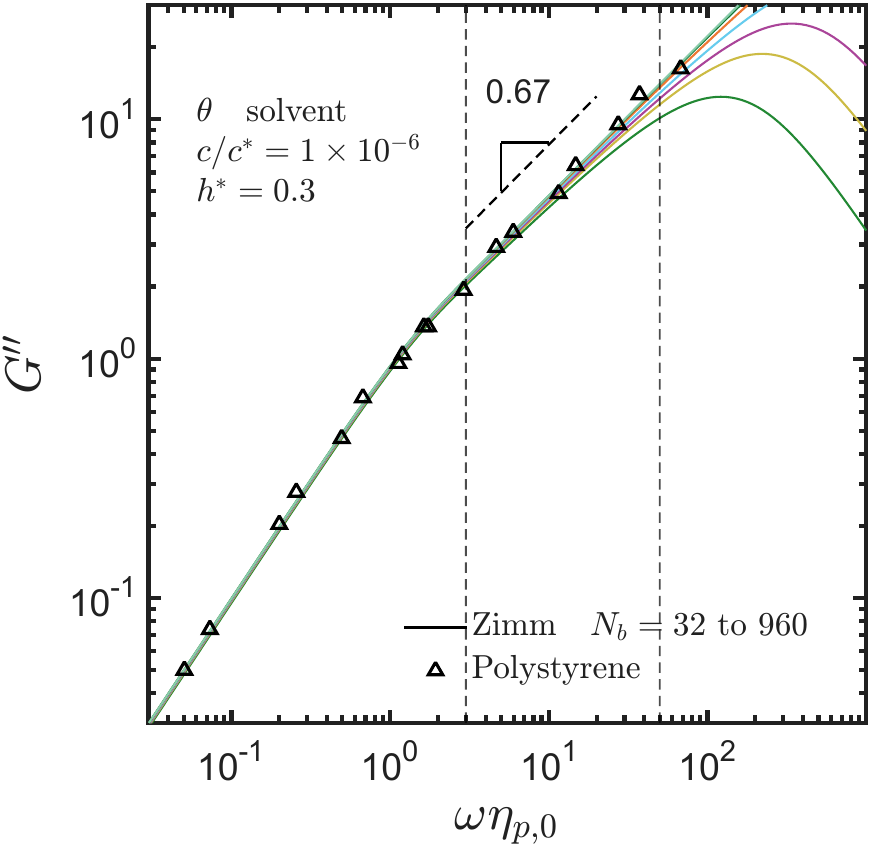} \\
             \large{(b)} 
\end{tabular}
	\end{center}
	\vskip-15pt
	\caption{Finite-chain effects in the loss modulus $G''(\omega)$ under $\theta$ solvent conditions in the dilute limit. Comparison of $G''(\omega)$ from the Zimm analytical model with different chain discretizations ($N_b = 32$ to $960$) \citep{Bird1987} at hydrodynamic interaction parameter (a) $h^* = 0.2$ and (b) $h^* = 0.3$ with experimental data (triangles) for polystyrene from \citet{Johnson1970}. Increasing $N_b$ systematically extends the scaling regime to higher frequencies. Here, the dashed lines correspond to the frequency window where successive fine graining is performed. The slope represents the Zimm power law exponent for a $\theta$ solvent.}
	\label{fig:finiteNb_rouse}
	\vskip-15pt	
\end{figure}

Early attempts to compare predictions by bead-spring chain models with these experimental observations were based on the Zimm model, which uses pre-averaged hydrodynamic interactions. In terms of the \emph{draining} parameter $h= h^* N_b^{1/2}$, which measures the strength of hydrodynamic interactions~\citep{Zimm1956, Ottinger1989b}, it was found that the Zimm model was excellent at accurately predicting data in $\theta$ solvents in the \emph{non-draining} limit $h \to \infty$~\citep{Johnson1970}. On the other hand, for good solvents, satisfactory fits to the data could be obtained with the Zimm model by adjusting the values of $h^*$ and $N_b$, suggesting that hydrodynamic interactions were \emph{partially-drained}~\citep{Johnson1970}. As pointed out by~\citet{LarsonBook}, who cites the thesis of Landry~\citep{Landry}, the rule-of-thumb for polystyrene for getting good fits to data with the Zimm model is to use bead-spring chains with one bead per roughly 5000--10,000 molecular weight, and a value of $h^* = 0.15$. For other polymer systems, the appropriate coarse-graining level depends on the polymer chemistry and molecular characteristics, such as the molecular weight of the Kuhn step. As will be demonstrated here, however, parameter-free and quantitatively accurate predictions of $G'$ and $G''$ can be obtained for both $\theta$ and athermal solvents in the non-draining limit. 

The storage and loss moduli, $G'(\omega)$ and $G''(\omega)$, for flexible polymer chains under $\theta$ and athermal solvent conditions predicted by the current Brownian dynamics simulations are displayed in Fig.~\ref{fig:exp_dilute} for chains with $N_b = 96$. It is clear that over the entire range of frequencies that have been probed, the predicted storage modulus $G'(\omega)$ exhibits excellent agreement with experiment for both solvent conditions, capturing both the magnitude and the scaling behaviour without adjustable parameters. On the other hand, while the loss modulus $G''(\omega)$ also agrees closely with experimental data at low and intermediate frequencies, systematic deviations are observed at higher frequencies, with the simulated $G''(\omega)$ falling below the experimental values. These deviations do not reflect a failure of the underlying physical model, but instead originate from a numerical artefact associated with the discrete representation of the polymer chain. In particular, an accurate description of the high-frequency viscoelastic response requires a sufficiently fine discretization of the polymer chain, such that the number of internal relaxation modes is large. 

\begin{figure*}[!t]
	\centerline{
		\begin{tabular}{c c}
			\includegraphics[width=8.35cm,height=!]{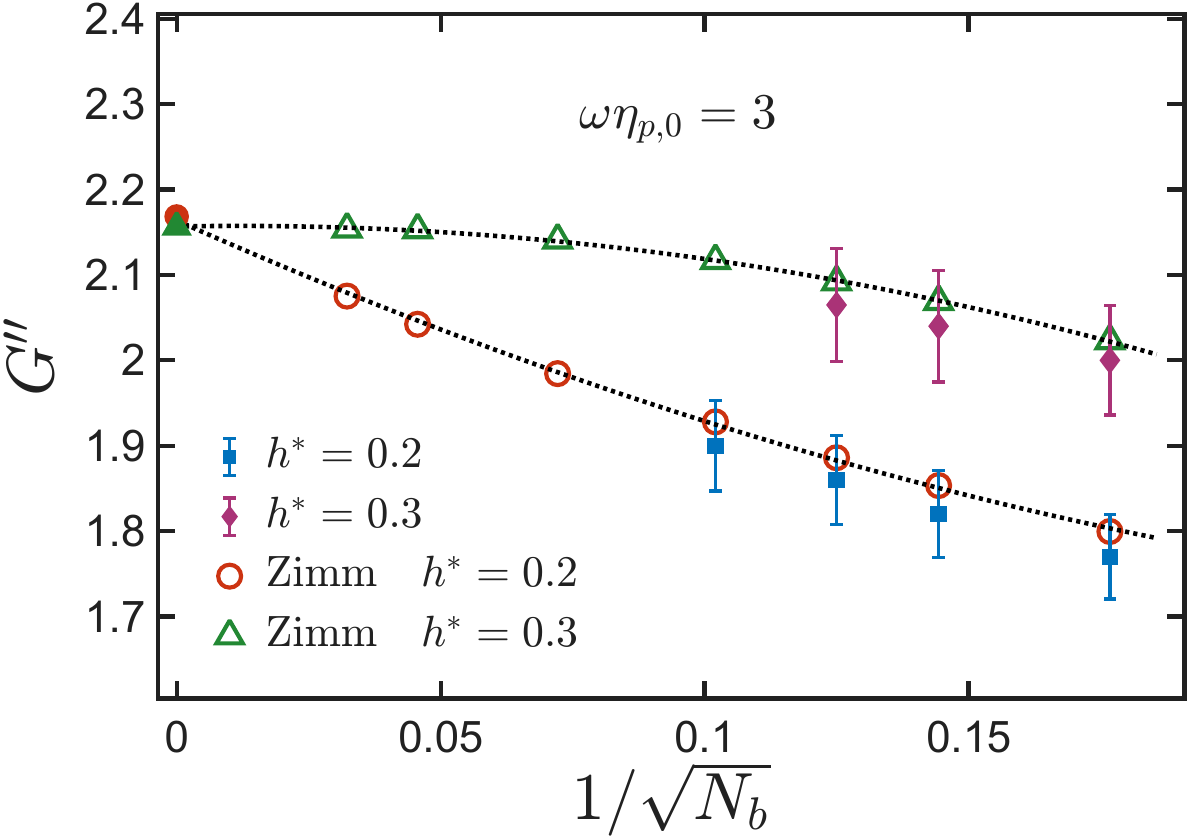} &
			\includegraphics[width=8.35cm,height=!]{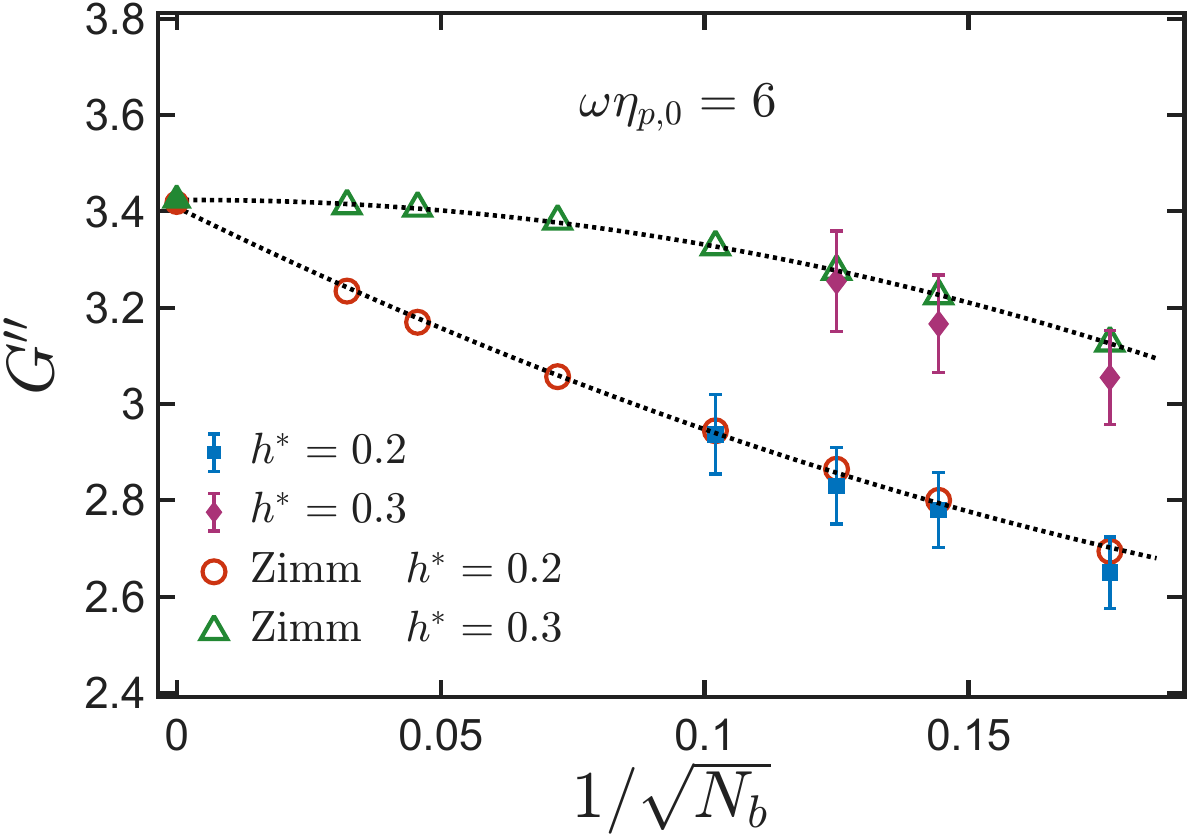} \\
			\large{(a)} &  \large{(b)} \\[2pt]        
			\includegraphics[width=8.35cm,height=!]{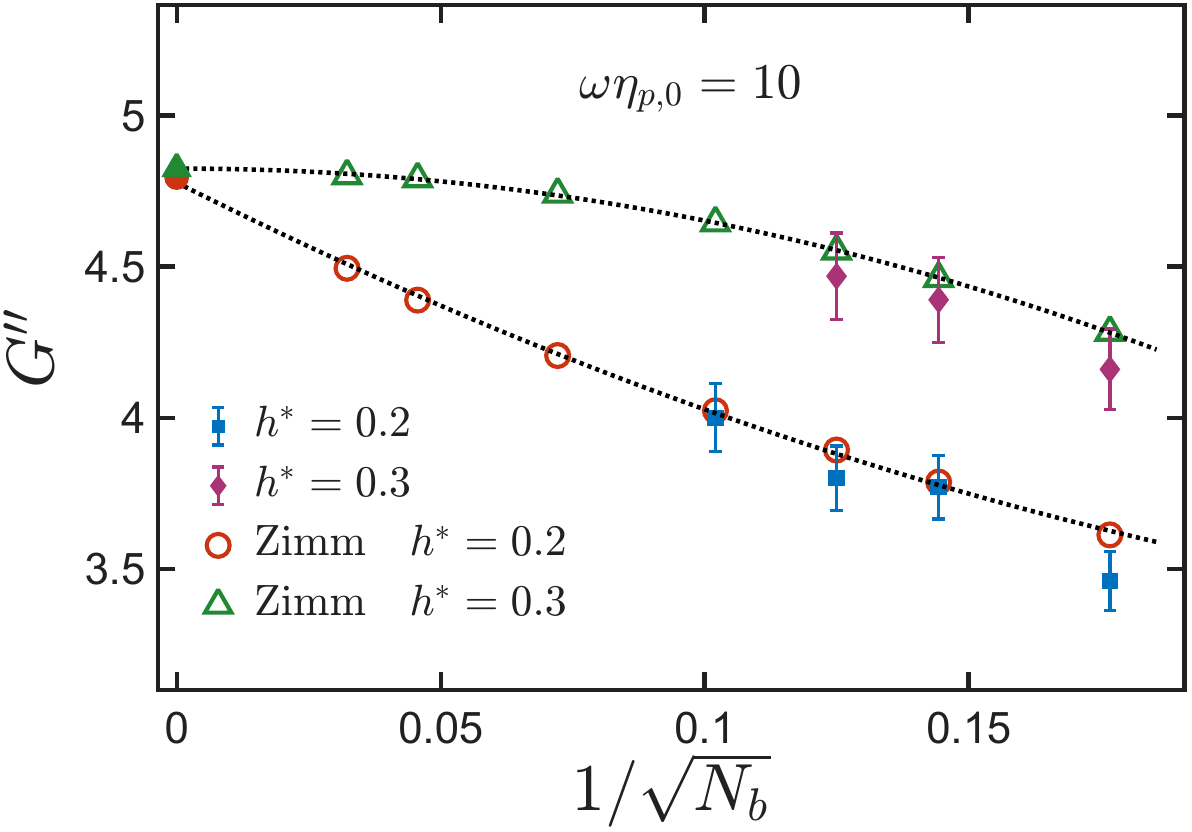} &
			\includegraphics[width=8cm,height=!]{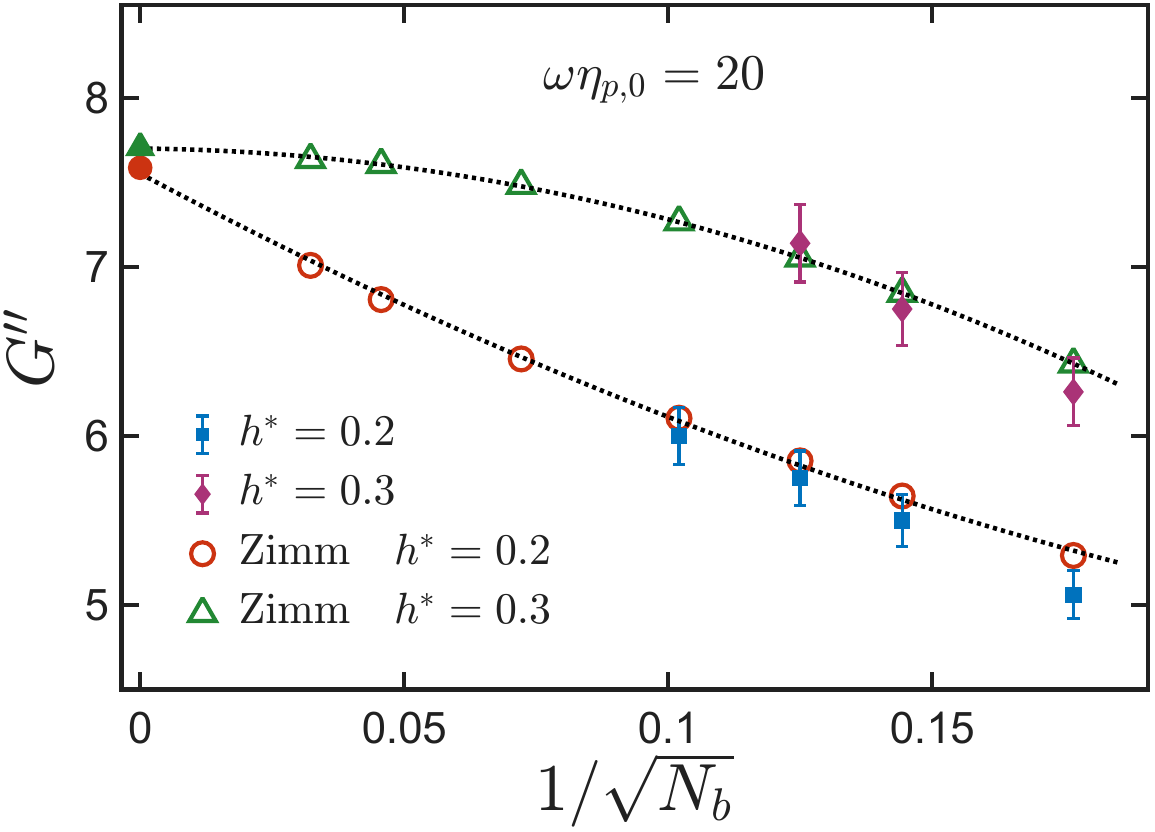} \\
			\large{(c)} &  \large{(d)} \\[2pt]    
			\includegraphics[width=8.35cm,height=!]{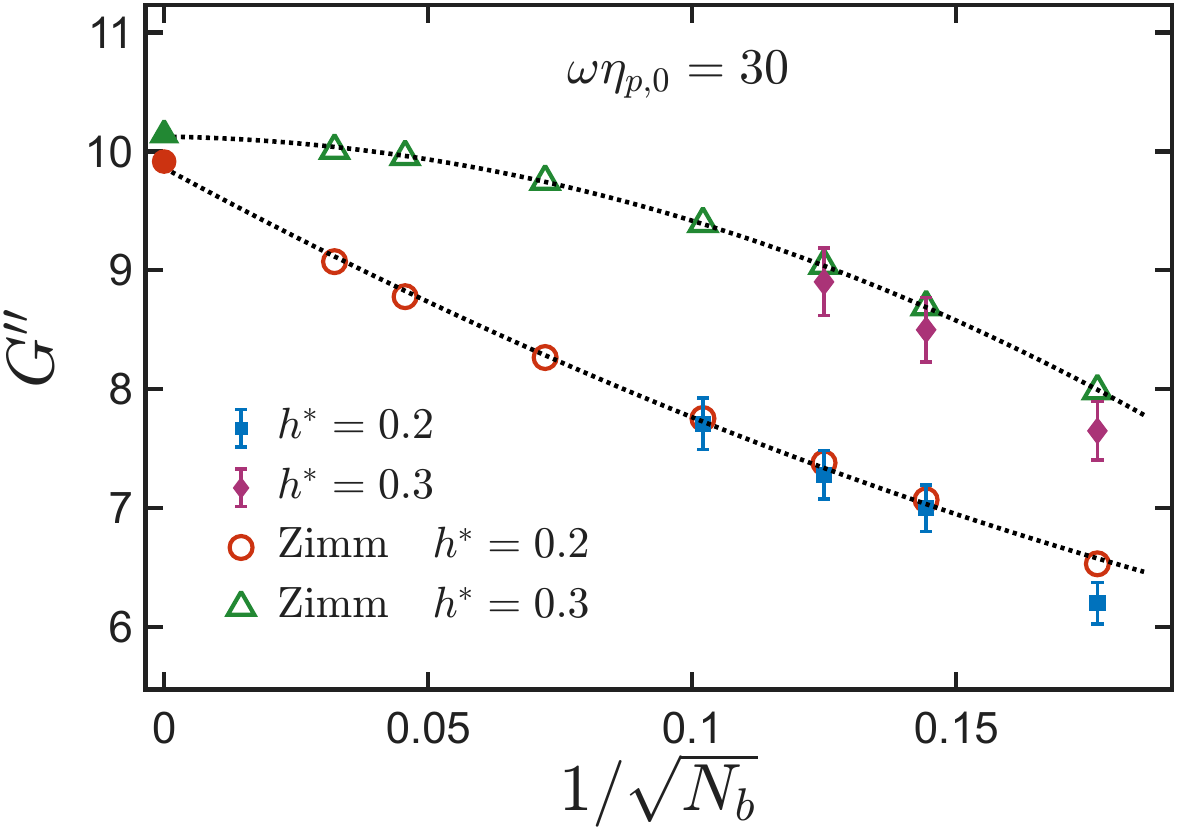} &
			\includegraphics[width=8.35cm,height=!]{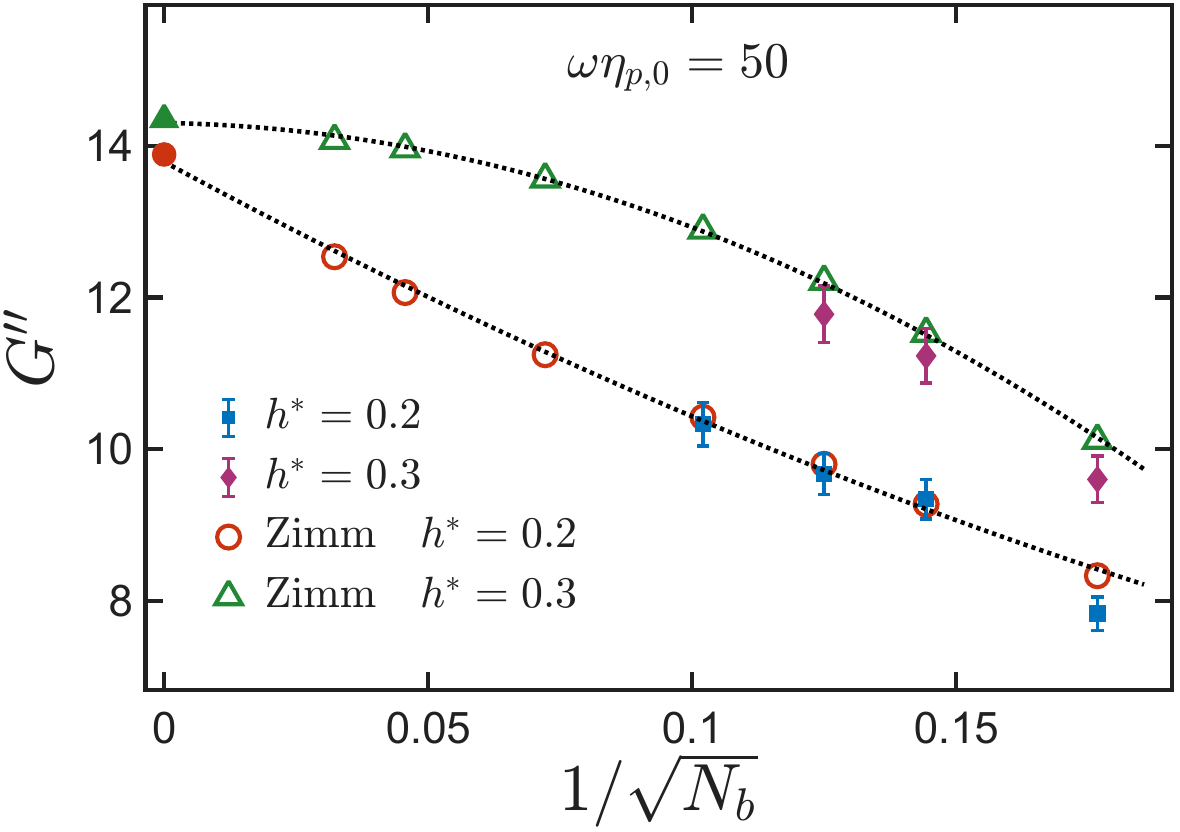} \\
			\large{(e)} &  \large{(f)}  
		\end{tabular}
	}
	\vskip-5pt
	\caption{Successive fine-graining extrapolation of finite chain data for the loss modulus $G''(\omega)$ to the infinite-chain length limit under $\theta$ solvent conditions, plotted as a function of $1/\sqrt{N_b}$. Each panel shows $G''(\omega)$ predicted by Zimm theory (hollow symbols) for varying chain discretizations, $N_b = 32$ to $960$, at $h^{*} = 0.2$ (hollow circles) and $h^{*} = 0.3$ (hollow triangles), together with the corresponding extrapolation to the $N_b \rightarrow \infty$ limit on the $y$-axis (filled symbols). Simulation results for bead-spring chains with $N_b = 32$ to $96$ in the infinitely dilute $\theta$ solvent limit are shown for $h^{*} = 0.2$ (square symbols) and $h^{*} = 0.3$ (diamond symbols). Panels correspond to reduced frequencies (a) $\omega \eta_{p,0} = 3$, (b) $\omega \eta_{p,0} = 6$, (c) $\omega \eta_{p,0} =10$, (d) $\omega \eta_{p,0} = 20$, (e) $\omega \eta_{p,0} = 30$, and (f) $\omega \eta_{p,0} = 50$.}
	\label{fig:sfg_examples}
	\vskip-15pt
\end{figure*}

To further elucidate the origin of the high-frequency deviations observed in the loss modulus, the dependence of $G''(\omega)$ on chain discretization was examined by comparing analytical Zimm predictions \citep{Bird1987} for a wide range of bead numbers. Fig.~\ref{fig:finiteNb_rouse} shows $G''(\omega)$ for $\theta$ solvent conditions at infinite dilution, with $N_b$ ranging from $32$ to $960$, for hydrodynamic interaction parameters $h^* = 0.2$ and $0.3$, together with experimental data. As the chain is progressively fine-grained by increasing $N_b$, which leads to an increase in the number of internal modes, the position of the maximum in $G''(\omega)$ shifts to higher frequencies, and the scaling regime extends over a broader range. This behavior reflects the fact that finer chain discretization introduces additional internal relaxation modes, thereby shifting the truncation of the relaxation spectrum to higher frequencies, clearly demonstrating that the observed deviations in $G''(\omega)$ arise from finite discretization rather than from missing physics.

The concept of successive fine-graining (SFG) provides a systematic framework for obtaining parameter-free predictions from coarse-grained bead-spring chain models by exploiting the universal behavior of long-chain polymer solutions. The underlying idea is based on the observation that, in the long chain length limit, polymer properties become independent of microscopic details and the specific choice of model parameters, and depend only on macroscopic variables such as solvent quality and concentration. In this approach, data are first obtained for finite chain discretizations and subsequently extrapolated to the long chain length limit, $N_b \to \infty$, in order to recover this universal behavior. At equilibrium, this technique has been widely used to obtain universal predictions from analytical theories and molecular simulations~\citep{Sunthar2006,Prakash1999,Jain2012,Garcia1984,Freire1986}.

\begin{figure}[t]
	\centering
	\includegraphics[width=8.35cm,height=!]{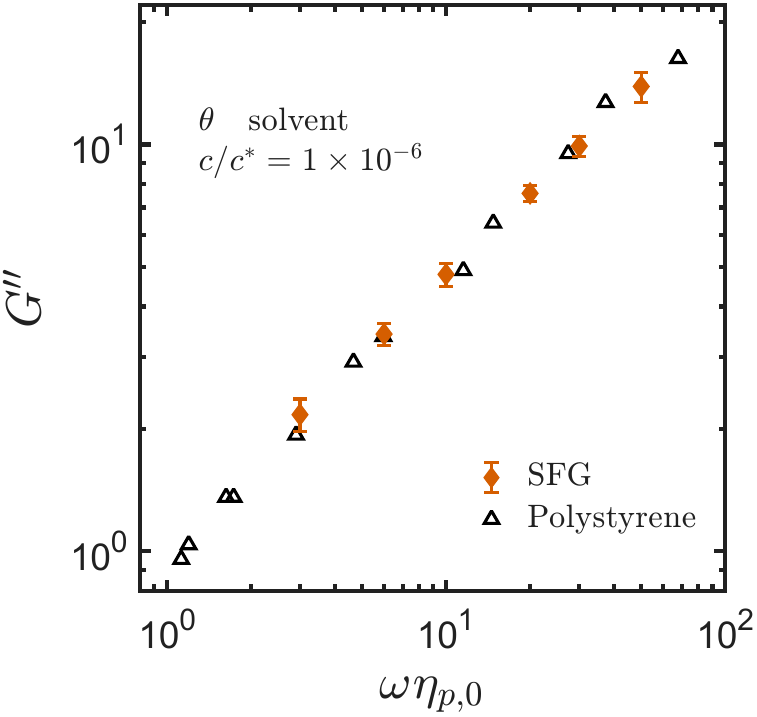}
	\vskip-10pt
	\caption{Loss modulus obtained by applying the successive fine-graining (SFG) procedure to finite chain values predicted by the analytical Zimm model to the infinite-chain length limit for different frequencies (filled diamonds), compared with experimental data (triangles). Error bars on the SFG data reflect the difference in extrapolated values for the two choices of the hydrodynamic interaction parameter, $h^* = 0.2$ and $0.3$.}
	\label{fig:finiteNb_rouse2}
	\vskip-15pt
\end{figure}

Fig.~\ref{fig:sfg_examples} illustrates representative examples of the successive fine-graining extrapolation procedure of the analytical Zimm predictions for the loss modulus $G''(\omega)$ at selected reduced frequencies $\omega \eta_{p,0} = 3$ to $50$. For each frequency, $G''(\omega)$ is plotted as a function of $1/\sqrt(N_b)$ for chain discretizations ranging from $N_b = 32$ to $960$, and for two values of the hydrodynamic interaction parameter, $h^* = 0.2$ and $0.3$. The square and diamond symbols represent simulation data obtained at finite values of $N_b$ for the two choices of $h^*$. Due to the limited range of available simulation data, the extrapolation is performed using only the analytical Zimm results. While the finite $N_b$ results show a clear dependence on $h^*$, the extrapolated values converge to a common limit as $N_b \to \infty$ across all frequencies. For infinitely long chains, the value of the draining parameter, $h = h^* N_b^{1/2} \to \infty$, corresponding to the non-draining limit. The convergence to a common value, independent of the choice of $h^*$, is consistent with the theoretical expectation that universal property predictions are obtained in the non-draining limit~\citep{Osaki,Jain2012,Kroger2000,Ottinger1987,Ottinger1989b,Prakash1997}.

\begin{figure*}[!ht]
	\centerline{
		\begin{tabular}{c c}
			\includegraphics[width=8.35cm,height=!]{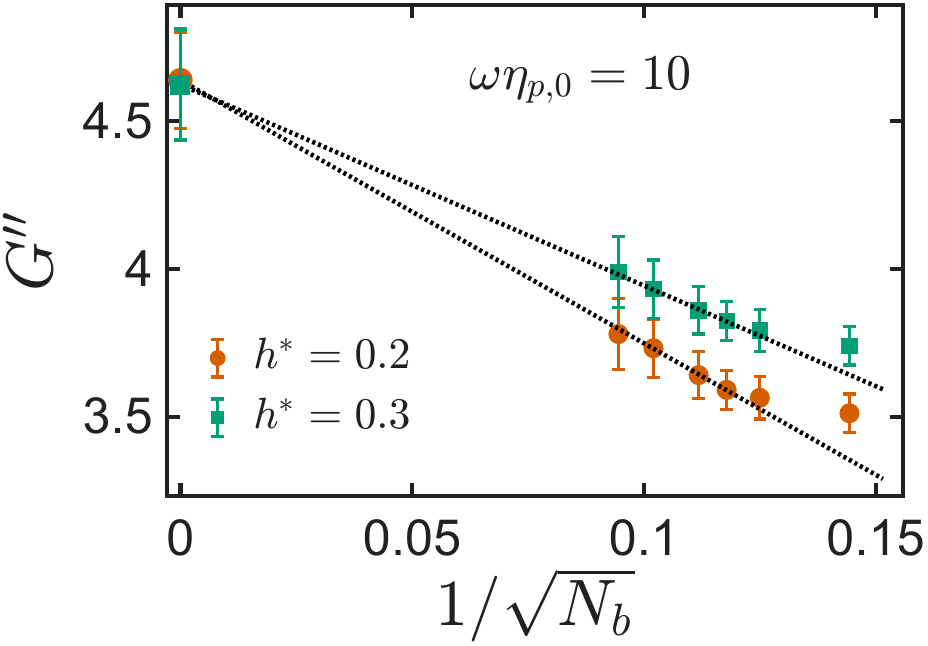} &
			\includegraphics[width=8.35cm,height=!]{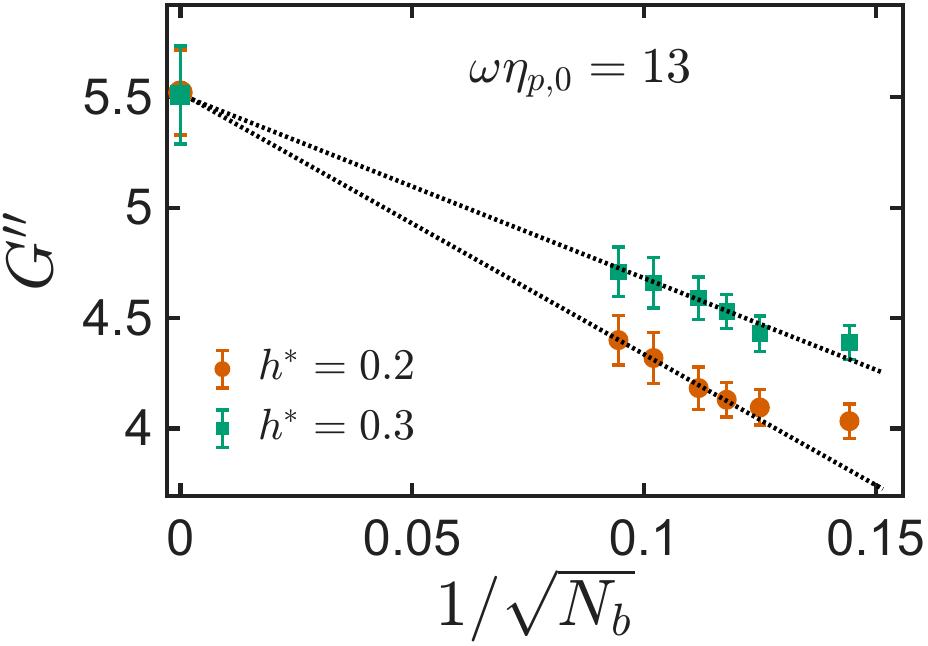} \\
			\large{(a)} &  \large{(b)} \\[5pt]          
			\includegraphics[width=8.35cm,height=!]{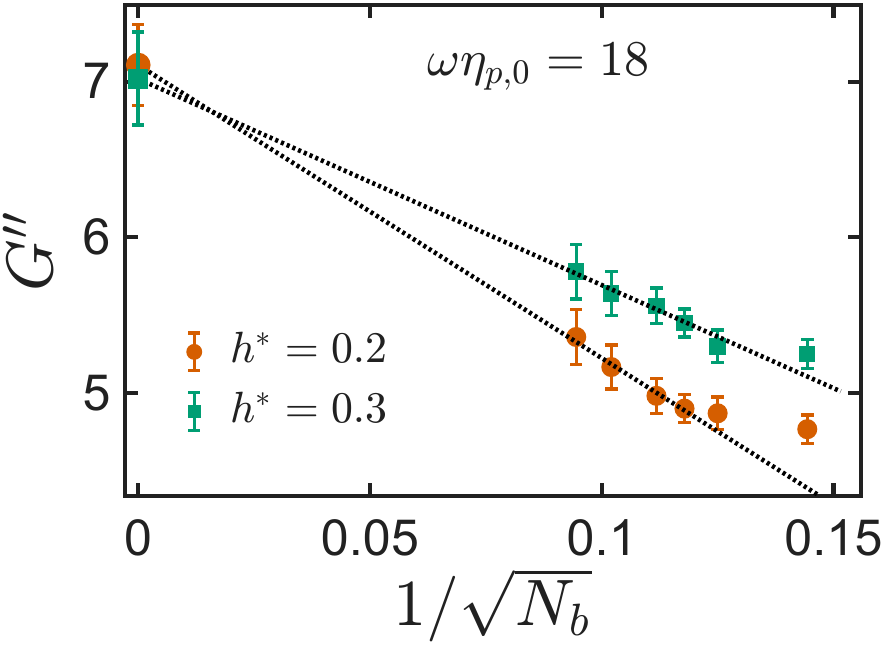} &
			\includegraphics[width=8.35cm,height=!]{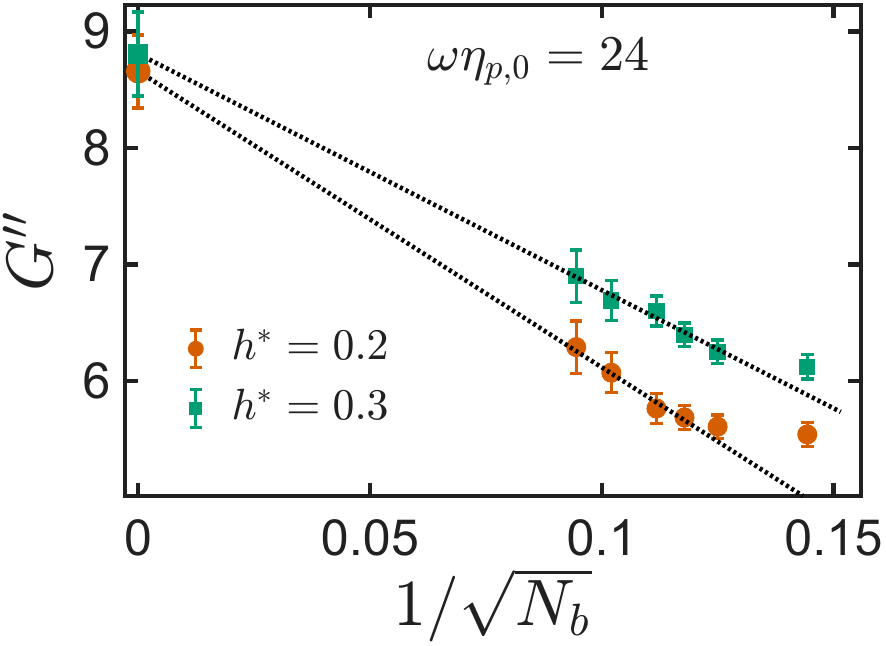} \\
			\large{(c)} &  \large{(d)}    
		\end{tabular}
	}
	\vskip-5pt
	\caption{
		Successive fine-graining extrapolation of finite size data for the loss modulus $G''(\omega)$ to the infinite-chain length limit under athermal solvent conditions, plotted as a function of $1/\sqrt{N_b}$. Simulation data are shown for $h^{*} = 0.2$ (circles) and $h^{*} = 0.3$ (squares), together with the corresponding extrapolation to the $N_b \rightarrow \infty$ limit (filled symbols). The dashed lines represent least-squares linear fits to the four largest chain lengths used in the extrapolation ($N_b=64$, $80$, $96$, and $112$). Panels correspond to reduced frequencies (a) $\omega \eta_{p,0} = 10$, (b) $\omega \eta_{p,0} = 13$, (c) $\omega \eta_{p,0} =18$, and (d) $\omega \eta_{p,0} = 24$. The error bars in the extrapolated values are estimated using a least-squares fitting procedure.
	}
	\label{fig:sfg_good}
	\vskip-15pt
\end{figure*}

\begin{figure}[t]
	\centering
	   \includegraphics[width=8.35cm,height=!]{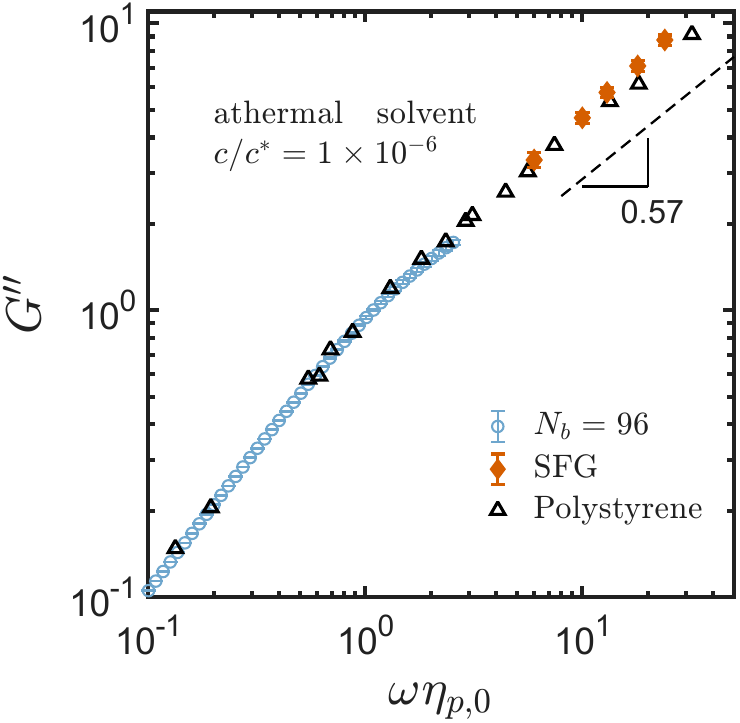}
	   \vskip-10pt
	\caption{Loss modulus $G''(\omega)$ for an athermal solvent. Open circles represent Brownian dynamics simulation results for a finite chain with $N_b = 96$, while filled diamonds correspond to values obtained using successive fine-graining (SFG) extrapolated to the infinite-chain length limit. Error bars on the SFG data reflect the variation in extrapolated values obtained using two choices of the hydrodynamic interaction parameter, $h^* = 0.2$ and $0.3$. The slope indicates the expected Zimm scaling exponent for athermal solvent conditions.}
	\label{fig:athermal_sfg}
	\vskip-15pt
\end{figure}

The use of $1/\sqrt{N_b}$ as the extrapolation variable is motivated by the scaling of the leading-order corrections to the long-chain limit \citep{Ottinger1987,Ottinger1989b,Prakash1997}. The loss moduli exhibit a nearly linear dependence on $1/\sqrt{N_b}$ for sufficiently large $N_b$, enabling reliable extrapolation to the infinite-chain length limit. Deviations from linear behavior are observed at smaller $N_b$, where higher-order corrections to the leading order scaling become significant. To account for these nonlinear finite-chain effects, the values of $G''(\omega)$ obtained at fixed frequency for different $N_b$ are extrapolated to $N_b \to \infty$ using a rational function extrapolation algorithm \citep{Press1992}.

All viscoelastic quantities presented in the preceding sections are normalized by the zero-shear rate viscosity $\eta_{p,0}$, which collapses the data across different chain discretizations onto a master curve. However, it is important to note that the absolute values of $\eta_{p,0}$ differ systematically between the analytical Zimm model and the Brownian dynamics simulations. The Zimm model employs a pre-averaged treatment of hydrodynamic interactions, whereas the simulations incorporate fluctuating hydrodynamic interactions.

The inclusion of hydrodynamic interaction fluctuations is known to produce quantitative differences in transport and rheological properties relative to predictions based on pre-averaging hydrodynamic interactions \citep{Kroger2000,Prakash1997}. The clearest analytical demonstration of this effect is provided by comparing the pre-averaged Zimm model with the Gaussian approximation \citep{Ottinger1989}, which incorporates fluctuating hydrodynamic interactions. The evolution equation for the covariance matrix in the Gaussian approximation contains additional terms arising from fluctuations in hydrodynamic interactions, and these terms are responsible for the observed differences in rheological properties, including the zero-shear rate viscosity \citep{Prakash1999b}.

For long chains, the Gaussian approximation predicts a ratio $R_0=\eta_{p,0}^{\mathrm{GA}}/\eta_{p,0}^{\mathrm{Zimm}}$ in the range 0.72--0.73 \citep{Ottinger1989}. Brownian dynamics simulations that explicitly account for fluctuating hydrodynamic interactions yield somewhat larger values, with $R_0 \approx 0.92$ reported by \citet{Kroger2000} and $R_0 \approx 0.89$ reported by \citet{Pan2014} in the long-chain limit. The values of $R_0$ reported in Table~1 for $\theta$-solvent conditions are consistent with the trend reported in these earlier Brownian dynamics studies.

Physically, hydrodynamic interactions alter the drag experienced by a polymer chain in solution. The inclusion of fluctuations in hydrodynamic interactions modifies this hydrodynamic drag relative to the pre-averaged description. Consequently, the stress-relaxation dynamics and the resulting viscous response differ quantitatively between the two treatments, leading to differences between the values of $\eta_{p,0}$ predicted by models accounting for fluctuating hydrodynamic interactions and the pre-averaged Zimm theory.

To recover the long chain length limit, the loss modulus obtained from the analytical Zimm model is extrapolated to $N_b \to \infty$ at fixed frequencies using the successive fine-graining procedure. The resulting $G''(\omega)$, shown in Fig.~\ref{fig:finiteNb_rouse2}, corresponds entirely to the extrapolated analytical Zimm predictions and not to simulation data. The extrapolated response exhibits excellent agreement with experimental data over the intermediate to high frequency range  where finite-chain effects are most significant, confirming that deviations at finite $N_b$ arise solely from discretization effects and can be systematically eliminated.

Fig.~\ref{fig:sfg_good} shows the successive fine-graining extrapolation of the loss modulus $G''$ for athermal solvents at $h^* = 0.2$ and $0.3$, plotted as a function of $1/\sqrt{N_b}$. In both cases, $G''$ exhibits an approximately linear dependence on $1/\sqrt{N_b}$ at large $N_b$, which justifies the use of linear extrapolation to estimate the asymptotic $N_b \rightarrow \infty$ limit. Due to the computational cost associated with simulations at very large chain lengths, extrapolations are now performed using the four largest chain lengths available ($N_b=64$, $80$, $96$, and $112$) while the smaller chain lengths ($N_b=32$ and $48$), where higher-order finite-chain correction effects become significant, are excluded from the fitting procedure. Although more sophisticated extrapolation schemes, such as rational function fits, may improve accuracy, the present dataset is insufficient to reliably constrain additional fitting parameters, and linear extrapolation therefore provides the most transparent estimate. The uncertainty in the extrapolated values is estimated using a least-squares fitting procedure \citep{Press1992} which incorporates the error in the simulation data, in the determination of the extrapolated values. Importantly, extrapolations performed independently for the two values of $h^*$ converge to the same limiting value within uncertainty, consistent with the expectation that the infinite-chain length limit is independent of the choice of $h^*$ and governed by the nondraining regime, thereby providing strong evidence for the robustness of the extrapolated results.
\begin{table}[htb!]
\caption{Extrapolated values of the loss modulus $G''$ obtained from the successive fine-graining procedure for dilute athermal-solvent solutions. The uncertainties are estimated using the least-squares fitting procedure described in the text.}
\label{tab:athermal_sfg}
\centering
\renewcommand{\arraystretch}{1.3}
\begin{tabular}{cc}
\hline
$\omega\lambda_H$ & $G''$ \\
\hline
6  & $3.34 \pm 0.18$ \\
10 & $4.58 \pm 0.23$ \\
13 & $5.51 \pm 0.27$ \\
18 & $7.11 \pm 0.33$ \\
24 & $8.66 \pm 0.42$ \\
\hline
\end{tabular}
\vskip-15pt
\end{table}

In Fig.~\ref{fig:athermal_sfg}, Brownian dynamics simulation predictions for athermal solvents, for finite-chains with $N_b = 96$ (where finite chains are sufficient to obtain accurate predictions at low to intermediate frequencies), are plotted simultaneously with the results of the SFG predictions obtained for intermediate to high frequencies (see Fig.~\ref{fig:sfg_good}). The extrapolated values of the loss modulus $G''$ together with their associated uncertainties are reported in Table~\ref{tab:athermal_sfg}. The composite simulation data is in excellent agreement with experimental results over the entire frequency regime, and the SFG results recover the expected Zimm scaling behavior for athermal solvent at high frequencies. It is clear that there is no necessity to make the partial-draining approximation in order to obtain quantitatively accurate predictions of experimental observations in the case of good solvents.

\begin{figure*}[!t]
\begin{center}
    \begin{tabular}{ccc}
        \includegraphics[width=5.6cm,height=!]{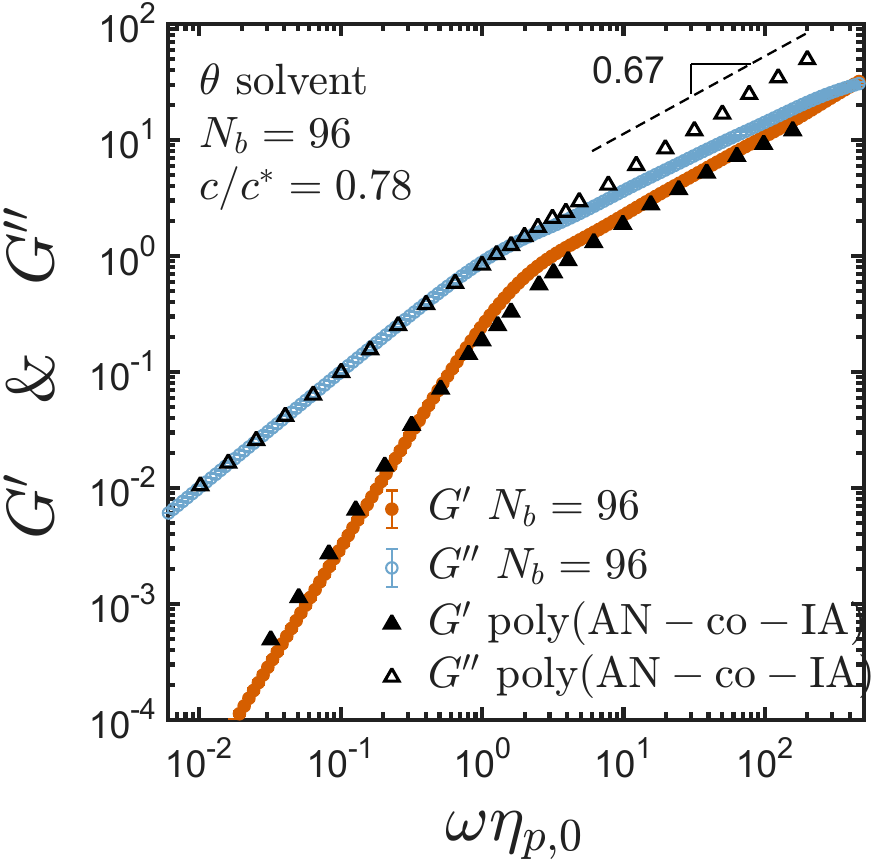} &
       \includegraphics[width=5.6cm,height=!]{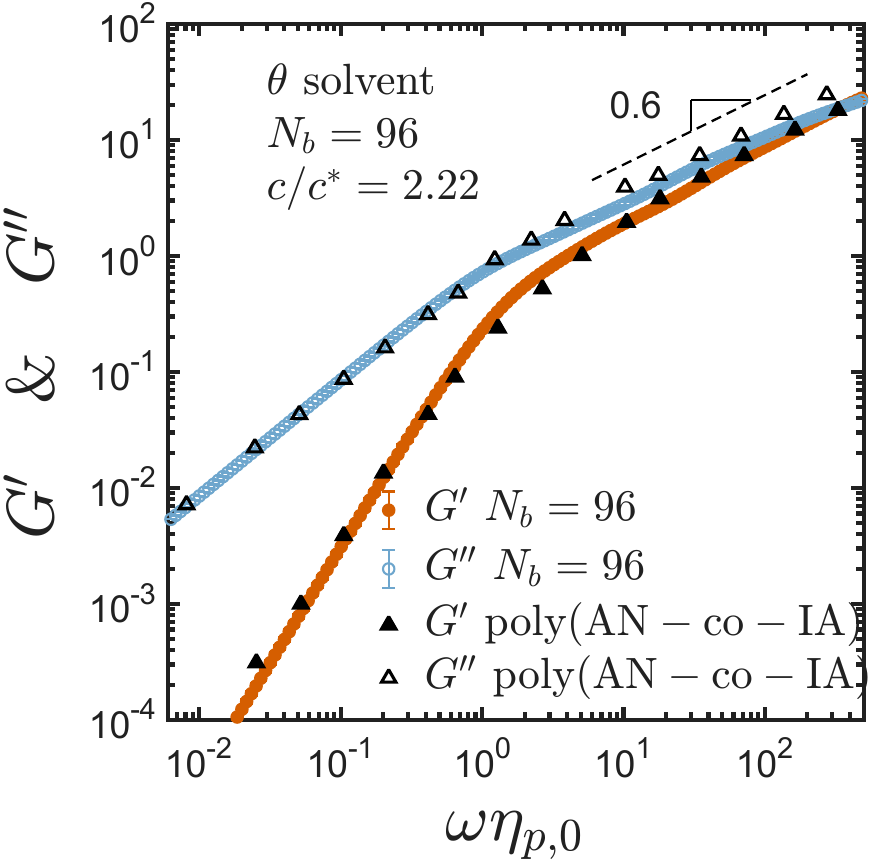}  &
   \includegraphics[width=5.6cm,height=!]{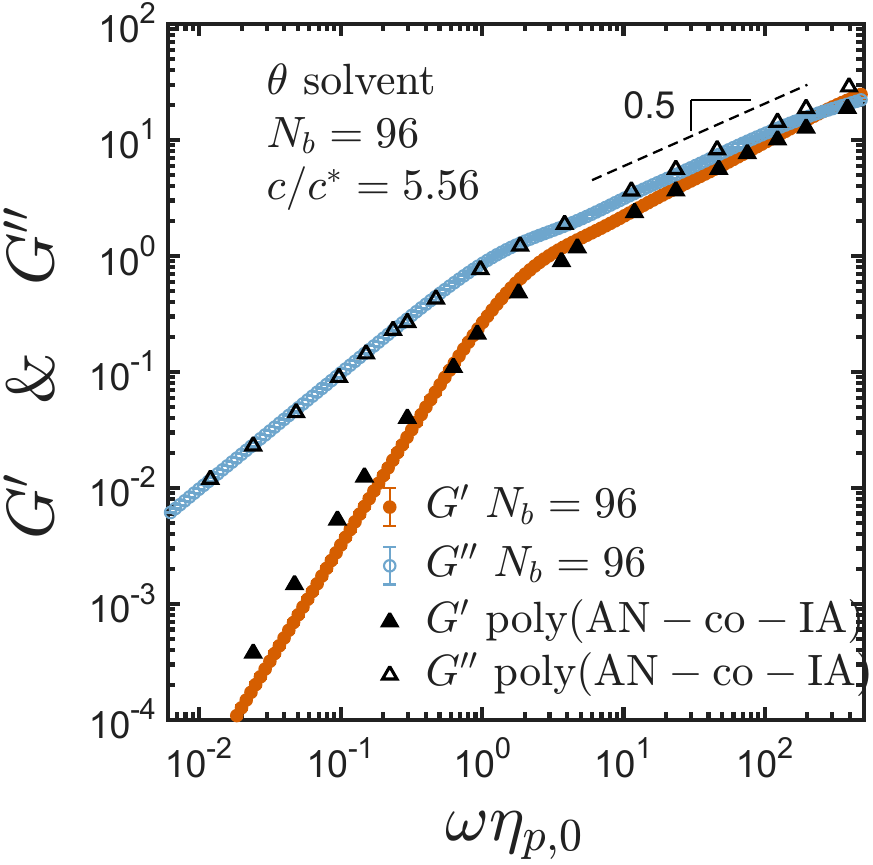} \\
        \large{(a)} &  \large{(b)} & \large{(c)} \\[5pt] 
          \end{tabular}
\end{center}
\vskip-15pt
	\caption{Comparison of dynamic moduli with the experimental data for poly(AN-co-IA) at concentrations: (a) $c/c^{*} = 0.78$, (b) $c/c^{*} = 2.22$, (c) $c/c^{*} = 5.56$ in a $\theta$ solvent. The non-dimensional dynamic moduli, $G'(\omega)$ (filled) and $G''(\omega)$ (hollow) are plotted as a function of frequency scaled with zero-shear viscosity $\eta_{p,0}$. Experimental data for poly(AN-co-IA) is taken from \citet{Zhu2012}. Simulation results correspond to chains of discretization $N_b=96$. Here the slopes represent the power law exponents calculated from the high frequency experimental data.}
	\label{fig:exp_semidilute}
\vskip-10pt
\end{figure*}
\begin{figure*}[tbhp]
    \centerline{
    \begin{tabular}{c c}
        \includegraphics[width=8.15cm,height=!]{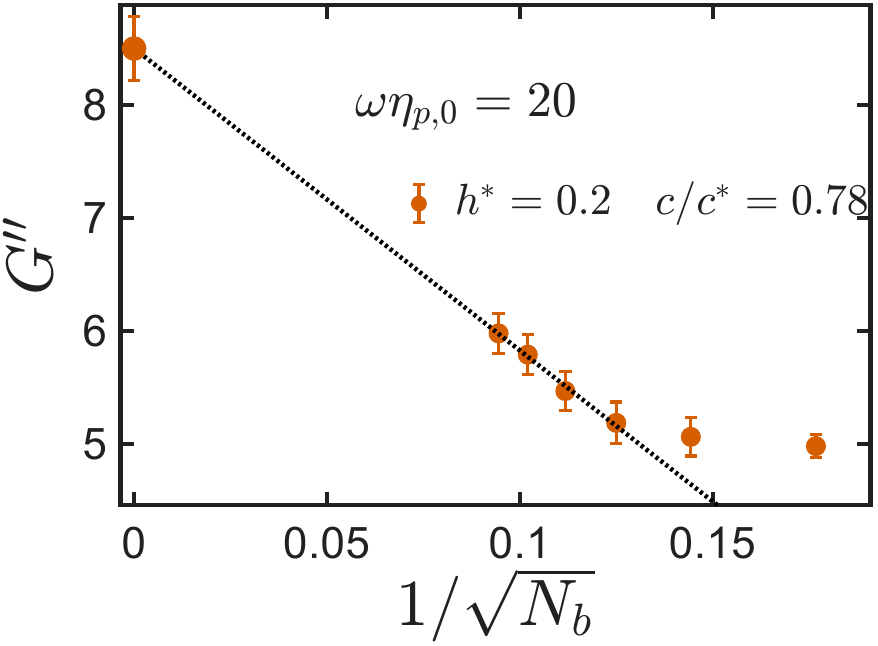} &
        \includegraphics[width=8.25cm,height=!]{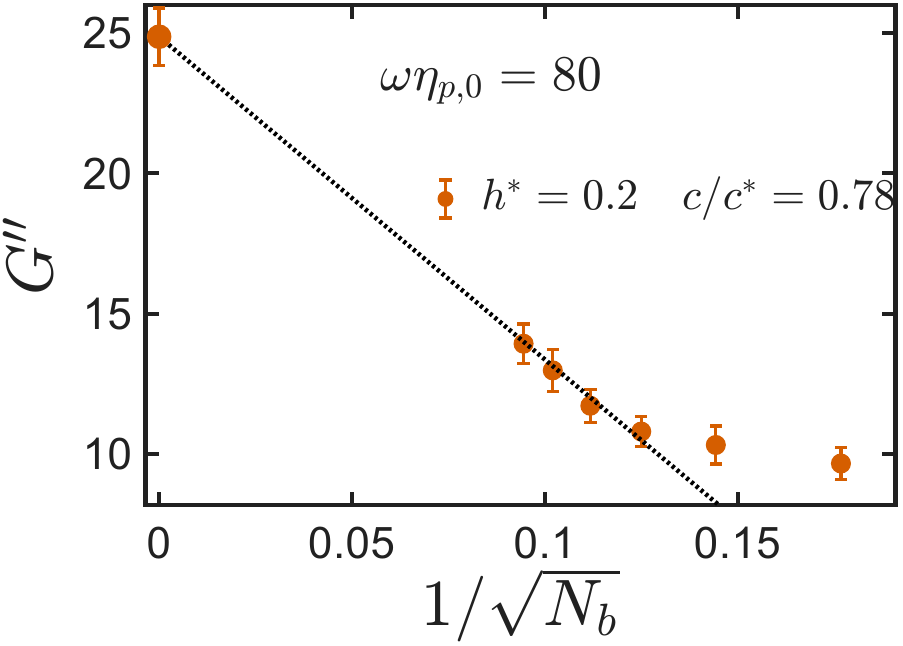} \\[10pt] 
        \large{(a)} &  \large{(b)} \\[10pt]        
        \includegraphics[width=8.25cm,height=!]{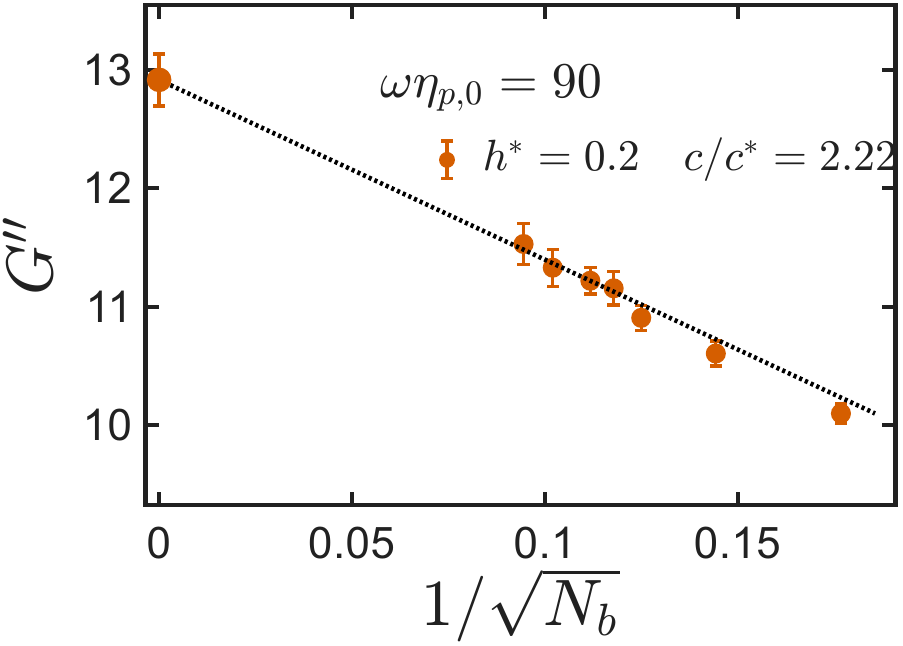} &
\includegraphics[width=8.25cm,height=!]{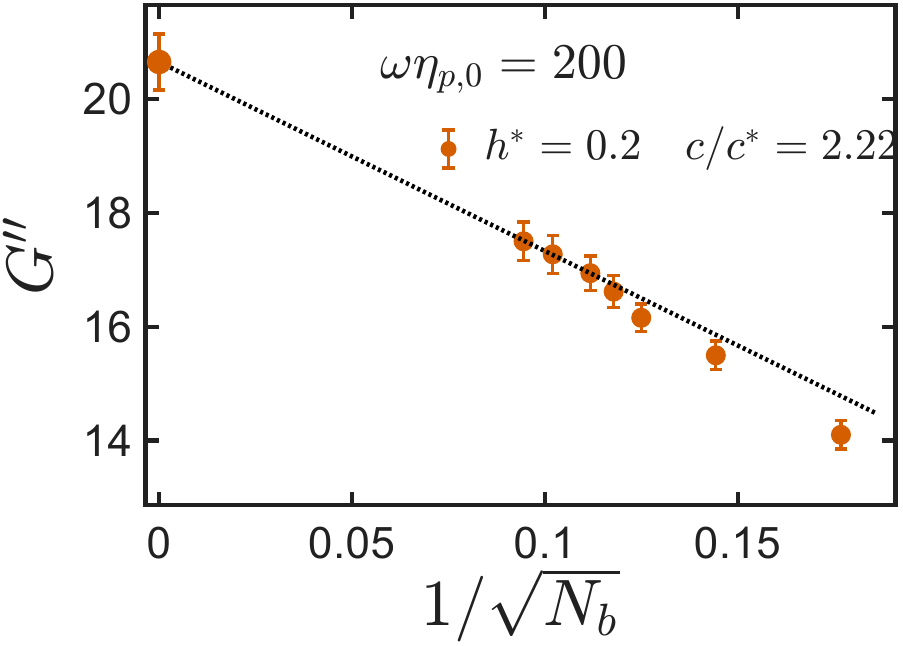} \\[10pt] 
      \large{(c)} &  \large{(d)} \\[10pt]    
\includegraphics[width=8.25cm,height=!]{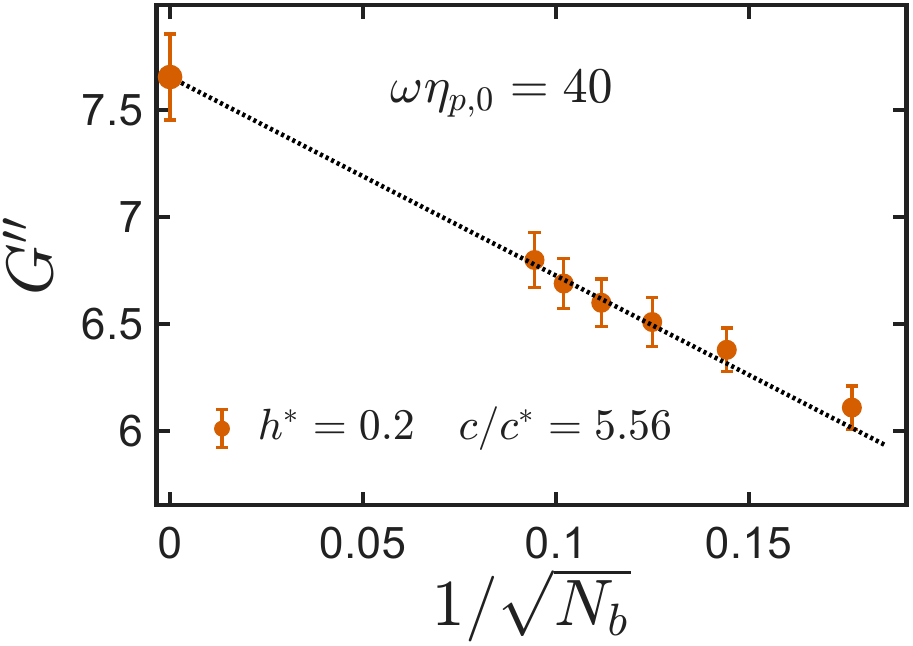} &
\includegraphics[width=8.25cm,height=!]{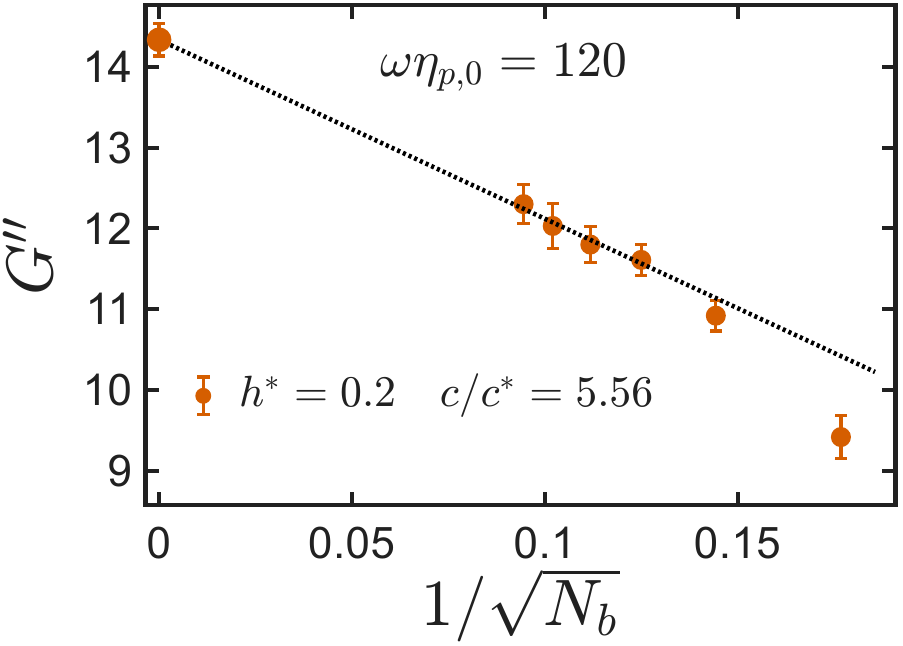} \\[10pt] 
          \large{(e)} &  \large{(f)}  
    \end{tabular}
    }
	\caption{
		Successive fine-graining extrapolation of finite chain data for the loss modulus $G''(\omega)$ to the infinite-chain length limit in semidilute solutions, plotted as a function of $1/\sqrt{N_b}$ for three different concentrations: (a) \& (b) $c/c^* = 0.78$, (c) \& (d) $c/c^* = 2.22$, and (e) \& (f) $c/c^* = 5.56$. In each panel, $G''(\omega)$ is shown for varying chain discretizations, $N_b = 32$ to $112$, at $h^* = 0.2$ (circle symbols), together with the corresponding extrapolation to the $N_b \rightarrow \infty$ limit (solid symbols). The dashed lines represent least-squares linear fits to the four largest chain lengths used in the extrapolation ($N_b=64$, $80$, $96$, and $112$).  }
	\label{fig:sfg_semidilute_combined}
\end{figure*}
\begin{figure*}[h!t]
\begin{center}
\begin{tabular}{c}
        \includegraphics[width=7.3cm,height=!]{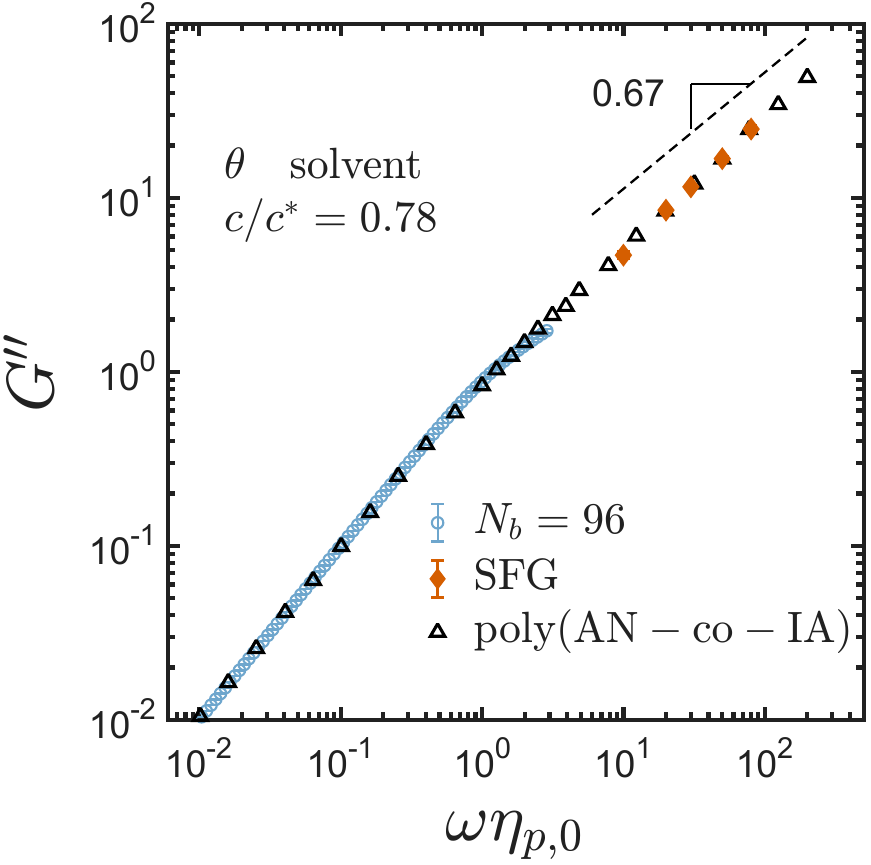} \\
          \large{(a)}\\[1pt] 
        \includegraphics[width=7.3cm,height=!]{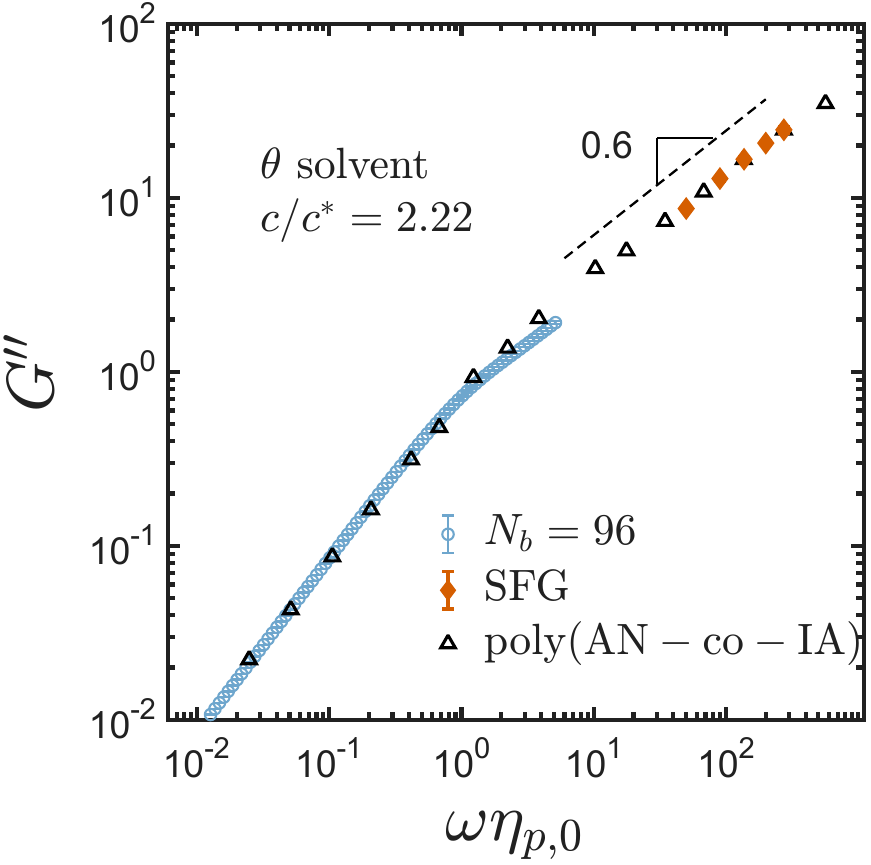} \\
             \large{(b)} \\[1pt]
               \includegraphics[width=7.3cm,height=!]{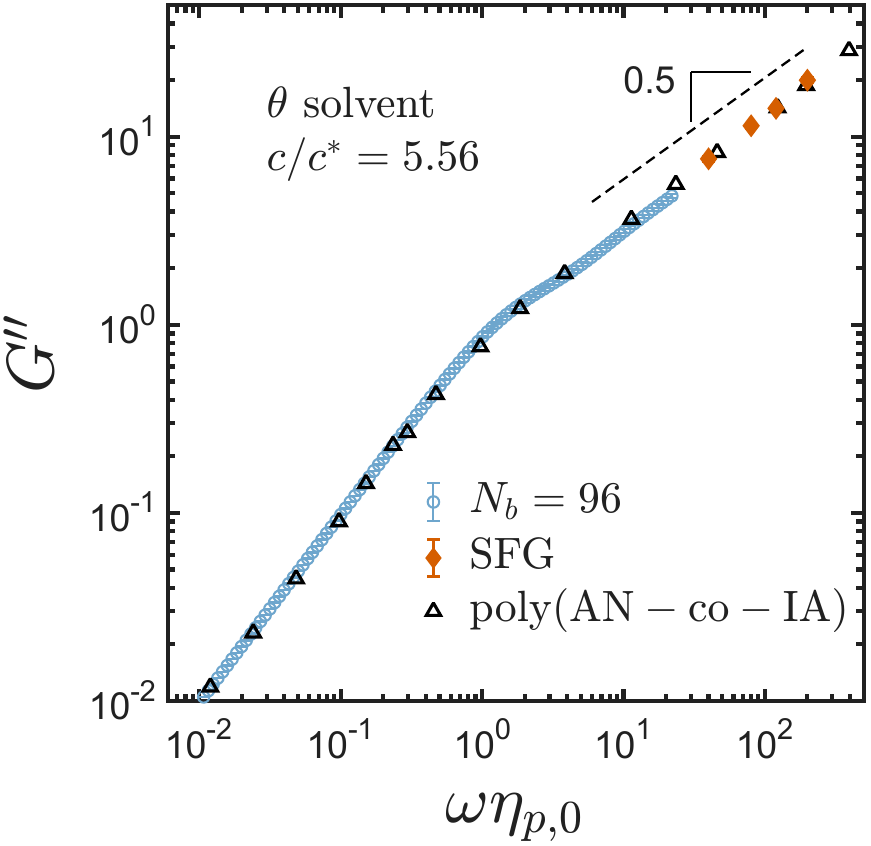} \\
             \large{(c)} \\[1pt]
\end{tabular}
\end{center}
\vskip-18pt
	\caption{
		Comparison of the loss modulus $G''(\omega)$ in semidilute solutions at different concentrations: (a) $c/c^* = 0.78$, (b) $c/c^* = 2.22$, and (c) $c/c^* = 5.56$. Open circles represent Brownian dynamics simulation results for a finite chain with $N_b = 96$, filled diamonds correspond to successive fine-graining (SFG) predictions extrapolated to the infinite-chain length limit, and triangles denote experimental data for poly(AN-co-IA) \citep{Zhu2012}.
	}
	\label{fig:exp_comparison_semidilute}
    \vskip-10pt
\end{figure*}

Applying the SFG procedure leads to the prediction of the universal behaviour of polymer solutions in the long chain limit of the bead-spring chain model for a polymer. The validity of this procedure relies on the assumption that a polymer chain exhibits self-similar behaviour on all length scales. Within the context of blob theory, self-similarity in a polymer chain at equilibrium manifests itself through the presence of thermal blobs (that account for the temperature in terms of the solvent quality parameter $z$) and correlation blobs (that account for the concentration in terms of the variable $c/c^*$)~\cite{Prakash2019}. As a consequence, the conformations of polymer chains (in terms of thermal and correlation blobs) are identical in all polymer solutions, regardless of polymer-solvent chemistry, at the same values of $z$ and $c/c^*$. In reality, however, self-similarity ceases to be a valid assumption when experimental observations probe length scales that are sufficiently small that local chemical structure, backbone architecture and specific polymer--solvent interactions become distinguishable between different polymer solutions. A detailed discussion of the conditions under which the successive fine-graining procedure remains valid is given in~\citet{Sasmal2017} in the context of extensional flows, where the presence of Pincus blobs (that account for the Weissenberg number $W\!i$), in addition to thermal and correlation blobs, enters the consideration of self-similarity in polymer conformations. By examining the interplay between thermal, correlation and Pincus blobs for various concentration regimes, \citet{Sasmal2017} determine the value of $W\!i$ at which the size of the Pincus blob becomes of order of a Kuhn step, such that the flow ``penetrates" to length scales below which the assumption of self-similarity is no longer valid, and the SFG procedure fails. In the context of the present application of the SFG procedure to the near-equilibrium linear viscoelastic behaviour of polymer solutions, one can anticipate that it breaks down at sufficiently high frequencies $\omega^*$, such that $(\omega^*)^{-1} \sim \mathcal{O} (\lambda_0)$, where $\lambda_0$ is the relaxation time of a Kuhn step. Indeed, a number of experimental, theoretical and molecular dynamics studies~\citep{Morris1988,Lodge1993, Larson2004,Muller1996,Saha2013,Schulz2015,Daldrop2018,Dalton2024} have discussed the non-universal linear viscoelastic behaviour of polymer solutions at high frequencies.

Larson \citep{Larson2004} proposed that torsional barriers associated with backbone bond rotations introduce a polymer-dependent cut-off in the relaxation spectrum, thereby limiting the universal behaviour predicted by bead--spring models. Using the molecular weight ($M=8.6\times10^5~\mathrm{g\,mol^{-1}}$) and intrinsic viscosity ($[\eta]=183~\mathrm{cm^3\,g^{-1}}$) reported by \citet{Johnson1970}, together with the solvent viscosity ($\eta_s\approx2~\mathrm{P}$ at $25^\circ\mathrm{C}$) and the estimated barrier-crossing time ($\tau_b\approx10^{-7}~\mathrm{s}$) reported by \citet{Larson2004}, this transition is estimated to correspond to a reduced frequency of $\omega_r\approx5\times10^4$, which is approximately 2.5 orders of magnitude higher than the maximum reduced frequency of $\omega_r\approx10^2$ explored by the experimental data and the SFG extrapolations shown in Fig.~8. More recent molecular dynamics studies \citep{Schulz2015,Daldrop2018,Dalton2024} have demonstrated the existence of additional dissipative mechanisms at high frequencies, commonly referred to as internal friction, arising from the coupling of conformational changes to other internal molecular motions, such as neighbouring bond rotations and torsional rearrangements.

In parallel, experimental studies by Lodge and co-workers \citep{Morris1988,Lodge1993} speculated that the surrounding solvent also departs from the continuum assumption at molecular length scales, This was further shown with atomistic molecular dynamics simulations by \citet{Muller1996} and \citet{Saha2013}. Polymer chains modify the rotational and translational dynamics of neighbouring solvent molecules, producing an effective local solvent viscosity that differs from the bulk solvent viscosity. These polymer-induced modifications of the local solvent friction provide an additional source of polymer-specific high-frequency behaviour that is absent from Brownian dynamics simulations employing a continuum solvent description. The agreement between the present simulation predictions and the experimental measurements over the frequency range considered here suggests that these molecular-scale effects are not significant for the systems investigated in this work.

\begin{table*}[t]
	\caption{Extrapolated values of the loss modulus $G''$ obtained from the successive fine-graining procedure for semidilute $\theta$-solvent solutions. The uncertainties are estimated using the least-squares fitting procedure described in the text.}
	\label{tab:semidilute_sfg}
	\centering
	\renewcommand{\arraystretch}{1.1}
	\begin{tabular}{cc|cc|cc}
		\hline
		\multicolumn{2}{c|}{$c/c^*=0.78$} &
		\multicolumn{2}{c|}{$c/c^*=2.22$} &
		\multicolumn{2}{c}{$c/c^*=5.56$} \\
		\hline
		$\omega\lambda_H$ & $G''$ &
		$\omega\lambda_H$ & $G''$ &
		$\omega\lambda_H$ & $G''$ \\
		\hline
		10  & $4.69 \pm 0.21$ &
		50  & $8.69 \pm 0.17$ &
		40  & $7.63 \pm 0.20$ \\
		
		20  & $8.50 \pm 0.28$ &
		90  & $12.92 \pm 0.22$ &
		80  & $11.43 \pm 0.29$ \\
		
		30  & $11.58 \pm 0.40$ &
		137 & $16.67 \pm 0.36$ &
		120 & $14.14 \pm 0.32$ \\
		
		50  & $16.81 \pm 0.60$ &
		200 & $20.65 \pm 0.49$ &
		200 & $19.97 \pm 0.42$ \\
		
		80  & $24.87 \pm 1.02$ &
		274 & $24.63 \pm 0.72$ &
		--- & --- \\
		\hline
	\end{tabular}
	\vskip-5pt
\end{table*}

\subsection{\label{sec:exp_semidil} Comparison with experiments: Semidilute solutions}

The simulated linear viscoelastic response in the semidilute unentangled regime is compared with experimental measurements reported by \citet{Zhu2012}. In that study, dynamic moduli were measured for poly(acrylonitrile-co-itaconic acid) also called poly(AN-co-IA), dissolved in the ionic liquid 1-butyl-3-methylimidazolium chloride, which behaves as a $\theta$ solvent for neutral polymer chains over the concentration range investigated. Fig.~\ref{fig:exp_semidilute} shows the storage and loss moduli, $G'(\omega)$ and $G''(\omega)$, for flexible polymer chains at concentrations $c/c^* = 0.78$, $2.22$, and $5.56$ under $\theta$ solvent conditions. The reduced frequency is defined using the experimentally measured polymer contribution to the zero-shear viscosity, $\eta_{p,0}$, through $\omega\eta_{p,0}M/(cN_Ak_BT)$. The corresponding values of $\eta_{p,0}$ and the molecular weight of the polymer are listed in Table~\ref{tab:exp_systems}. Simulation results are shown for chains of discretization $N_b = 96$ and are plotted using the same reduced units as in the preceding sections. Across all concentrations, $G'(\omega)$ exhibits excellent agreement with experimental data over the entire frequency range, capturing both the magnitude and the concentration-dependent evolution of the viscoelastic response as was observed previously in the case for dilute solutions (see Fig.~\ref{fig:exp_dilute}).

The loss modulus also shows good agreement with experiment at low and intermediate frequencies, with increasing concentration leading to a systematic shift toward Rouse-like behavior. At higher frequencies, deviations between simulation and experiment become apparent, similar to those observed in the dilute limit. However, with increase in $c/c^{*}$, $G^{''}$ compares better with the simulation data for higher frequencies. It is also observed that the separation between the storage and loss moduli decreases with increasing concentration. This behavior arises from the increased chain overlap in the semidilute regime, which lead to a stronger elastic response. As a result, the distinction between elastic and viscous contributions becomes less pronounced, leading to closer magnitudes of $G'(\omega)$ and $G''(\omega)$.
The simulation data suggests that the loss modulus approaches the intermediate scaling regime more rapidly with increasing concentration. This can be seen from the successive fine-graining plots at finite concentrations (Fig.~\ref{fig:sfg_semidilute_combined}), where the dependence of $G''(\omega)$ on $1/\sqrt{N_b}$ becomes increasingly linear over a wider range of $N_b$ as $c/c^*$ increases. In particular, at higher concentrations, the linear scaling is observed even for relatively small values of $N_b$, indicating that the system approaches the asymptotic scaling regime more quickly. This behavior is consistent with the emergence of more uniform, Rouse-like dynamics in semidilute solutions, where the effective relaxation spectrum is less sensitive to chain discretization.

Simulation data for low to intermediate frequencies obtained using finite-chains with $N_b = 96$, are plotted simultaneously with the results of SFG extrapolations for intermediate to high frequencies, across a range of concentrations in Fig.~\ref{fig:exp_comparison_semidilute}. The extrapolated values of the loss modulus $G''$ together with their associated uncertainties are reported in Table~\ref{tab:semidilute_sfg}. At all the concentrations considered, the composite simulation results exhibit excellent agreement with experimental data for  $G''(\omega)$ over a broad range of frequencies. In particular, SFG rectifies the systematic deviations  from experimental data observed at higher frequencies when finite-chain length simulations with $N_b = 96$ were carried out. The convergence of SFG predictions with experimental data across all concentrations highlights the ability of the methodology to recover the correct long chain length limit and provide parameter-free and quantitatively accurate predictions for semidilute polymer dynamics. Overall, the comparison demonstrates that the present simulations successfully capture the essential features of semidilute polymer linear viscoelasticity across a wide range of concentrations.

\section{\label{sec:conc} Conclusions}

In this work, the linear viscoelastic response of flexible polymer solutions in the dilute and semidilute unentangled concentration regimes is studied using Brownian dynamics simulations with excluded-volume and hydrodynamic interactions. By systematically varying concentration and chain discretization for $\theta$ and athermal solvents, and by analyzing both time- and frequency-domain linear viscoelastic functions, the present work provides a unified description of linear viscoelasticity for flexible polymer solutions. The main conclusions of this study are as follows:

\begin{itemize}
\item In the dilute limit, the stress relaxation modulus exhibits the expected Zimm-like power-law behavior, with solvent-quality-dependent exponents. When appropriately normalized with the zero-shear viscosity, data for different chain lengths collapse onto a universal master curve, confirming that the simulations correctly capture single-chain linear viscoelastic behavior.

\item In the semidilute unentangled regime, increasing concentration leads to a clear crossover from Zimm-like to Rouse-like behavior in both the time and frequency domains. This crossover reflects the progressive screening of hydrodynamic interactions as chain overlap increases, in agreement with the correlation-blob theory of semidilute polymer solutions.

\item Comparison with experimental measurements in both dilute and semidilute regimes shows excellent agreement for the storage modulus over the entire frequency range and for the loss modulus at low and intermediate frequencies. Deviations observed in the high-frequency loss modulus are traced to finite-chain discretization effects and do not affect the physically relevant scaling regimes.

\item By employing the successive fine-graining procedure and extrapolating finite chain data for the loss moduli data to the infinite-chain length limit, it is shown that the high-frequency behavior can also be effectively predicted. This demonstrates that finite-chain artefacts in the dynamic moduli can be controlled and eliminated, enabling parameter-free and quantitative comparison of dynamic moduli with experiments over an extended frequency range.

\item The excellent agreement with experiments and simulations shows that the dynamics of finite polymer solutions is currently well understood and can be completely captured by incorporating fluctuating excluded volume and hydrodynamic interactions into bead-spring chain models.
\end{itemize}

The results presented here provide a computationally tractable framework for understanding the linear viscoelastic properties of flexible polymer solutions across dilute and semidilute unentangled regimes. Future work will focus on extending this framework to semiflexible polymers in semidilute solutions, where chain stiffness introduces additional dynamical features and modifies the relaxation spectrum. Another important direction is to investigate intermediate solvent quality by considering finite values of the excluded volume potential well-depth $\epsilon$, enabling a systematic exploration of crossover behavior between $\theta$ and athermal solvent conditions. Additionally, it would be of interest to study the viscoelastic response of crosslinked polymer networks, where hydrodynamic screening, inter-chain correlations, and connectivity effects play a dominant role.

\begin{acknowledgments}
This research was undertaken with the assistance of resources from the National Computational Infrastructure (NCI Australia), an NCRIS enabled capability supported by the Australian Government. This work was also employed computational facilities provided by Monash University through the DUG, MASSIVE and MonARCH systems. We also acknowledge the funding and general support received from the IITB-Monash Research Academy.
\end{acknowledgments}
\bibliography{arxiv}

@article{Rouse1953,
    author = {Rouse, Prince E.},
    title = {{A} {T}heory of the {L}inear {V}iscoelastic {P}roperties of {D}ilute {S}olutions of {C}oiling {P}olymers},
    journal = {J. Chem. Phys.},
    volume = {21},
    number = {7},
    pages = {1272-1280},
    year = {1953},
    month = {07},
    issn = {0021-9606},
    doi = {10.1063/1.1699180},
    url = {https://doi.org/10.1063/1.1699180},
}

@article{Zimm1956,
    author = {Zimm, Bruno H.},
    title = {Dynamics of {P}olymer {M}olecules in {D}ilute {S}olution: {V}iscoelasticity, {F}low {B}irefringence and {D}ielectric {L}oss},
    journal = {J. Chem. Phys.},
    volume = {24},
    number = {2},
    pages = {269-278},
    year = {1956},
    month = {02},
    issn = {0021-9606},
    doi = {10.1063/1.1742462},
    url = {https://doi.org/10.1063/1.1742462},
}

@book{Rubinstein2003,
  title={Polymer {P}hysics},
  author={Rubinstein, Michael and Colby, Ralph H},
  year={2003},
  publisher={Oxford university press}
}

@book{Doi1988,
  title={The {T}heory of {P}olymer {D}ynamics},
  author={Doi, Masao and Edwards, Sam F and Edwards, Samuel Frederick},
  volume={73},
  year={1988},
  publisher={Oxford University Press}
}

@book{Bird1987,
  title={Dynamics of {P}olymeric {L}iquids, {V}olume 2: {K}inetic {T}heory},
  author={Bird, Robert Byron and Curtiss, Charles F and Armstrong, Robert C and Hassager, Ole},
  year={1987},
  publisher={Wiley}
}

@article{Adam1984,
	author = {Adam, M. and Delsanti, M.},
	title = {Viscosity and longest relaxation time of semi-dilute polymer solutions: {II}. {T}heta solvent},
	DOI= "10.1051/jphys:019840045090151300",
	url= "https://doi.org/10.1051/jphys:019840045090151300",
	journal = {J. Phys. France},
	year = 1984,
	volume = 45,
	number = 9,
	pages = "1513-1521",
}

@article{Ahlrichs2001,
  title = {Screening of hydrodynamic interactions in semidilute polymer solutions: {A} computer simulation study},
  author = {Ahlrichs, Patrick and Everaers, Ralf and D\"unweg, Burkhard},
  journal = {Phys. Rev. E},
  volume = {64},
  issue = {4},
  pages = {040501},
  numpages = {4},
  year = {2001},
  month = {Sep},
  publisher = {American Physical Society},
  doi = {10.1103/PhysRevE.64.040501},
  url = {https://link.aps.org/doi/10.1103/PhysRevE.64.040501}
}

@article{Chen2018,
    author = {Chen, Xun and Liang, Siwei and Wang, Shih-Wa and Colby, Ralph H.},
    title = {Linear viscoelastic response and steady shear viscosity of native cellulose in 1-ethyl-3-methylimidazolium methylphosphonate},
    journal = {J. Rheol.},
    volume = {62},
    number = {1},
    pages = {81-87},
    year = {2018},
    month = {01},
    issn = {0148-6055},
    doi = {10.1122/1.4990049},
    url = {https://doi.org/10.1122/1.4990049},
}

@article{Heo2005,
    author = {Heo, Youngsuk and Larson, Ronald G.},
    title = {The scaling of zero-shear viscosities of semidilute polymer solutions with concentration},
    journal = {J. Rheol.},
    volume = {49},
    number = {5},
    pages = {1117-1128},
    year = {2005},
    month = {09},
    issn = {0148-6055},
    doi = {10.1122/1.1993595},
    url = {https://doi.org/10.1122/1.1993595},
}

@article{Huang2010,
author = {Huang, Chien-Cheng and Winkler, Roland G. and Sutmann, Godehard and Gompper, Gerhard},
title = {Semidilute {P}olymer {S}olutions at Equilibrium and under {S}hear {F}low},
journal = {Macromolecules},
volume = {43},
number = {23},
pages = {10107-10116},
year = {2010},
doi = {10.1021/ma101836x},
URL = {https://doi.org/10.1021/ma101836x},
}

@article{Inoue2002,
author = {Inoue, Tadashi and Yamashita, Yasuhiro and Osaki, Kunihiro},
title = {Viscoelasticity of {P}olymers in $\theta$ {S}olvents around the {S}emidilute {R}egime},
journal = {Macromolecules},
volume = {35},
number = {24},
pages = {9169-9175},
year = {2002},
doi = {10.1021/ma020849v},
URL = {https://doi.org/10.1021/ma020849v},
}

@article{Johnson1970,
  title={Infinite-dilution viscoelastic properties of polystyrene in $\theta$-solvents and good solvents},
  author={Johnson, Robert M and Schrag, John L and Ferry, John D},
  journal={Polym. J.},
  volume={1},
  number={6},
  pages={742--749},
  year={1970},
  publisher={Nature Publishing Group}
}

@article{Pan2021,
  title={Universal solvent quality crossover of the zero shear rate viscosity of semidilute {DNA} solutions},
  author={Pan, Sharadwata and At Nguyen, Duc and Sridhar, Tamarapu and Sunthar, Papanasamoorthy and Ravi Prakash, J},
  journal={J. Rheol.},
  volume={58},
  number={2},
  pages={339--368},
  year={2014},
  publisher={AIP Publishing}
}

@article{Prakash2019,
  title={Universal dynamics of dilute and semidilute solutions of flexible linear polymers},
  author={Prakash, J Ravi},
  journal={Curr. Opin. Colloid Interface Sci.},
  volume={43},
  pages={63--79},
  year={2019},
  publisher={Elsevier}
}

@article{Tschoegl1964,
  title={Influence of hydrodynamic interaction on the viscoelastic behavior of dilute polymer solutions in good solvents},
  author={Tschoegl, NW},
  journal={J. Chem. Phys.},
  volume={40},
  number={2},
  pages={473--479},
  year={1964},
  publisher={AIP Publishing}
}

@article{Zhu2012,
  title={Linear viscoelasticity of poly (acrylonitrile-co-itaconic acid)/1-butyl-3-methylimidazolium chloride extended from dilute to concentrated solutions},
  author={Zhu, Xinjun and Chen, Xun and Saba, Hina and Zhang, Yumei and Wang, Huaping},
  journal={Eur. Polym. J.},
  volume={48},
  number={3},
  pages={597--603},
  year={2012},
  publisher={Elsevier}
}

@book{Ottinger2012,
  title={Stochastic processes in polymeric fluids: tools and examples for developing simulation algorithms},
  author={{\"O}ttinger, Hans C},
  year={2012},
  publisher={Springer Science \& Business Media}
}

@article{Soddemann2001,
  title={A generic computer model for amphiphilic systems},
  author={Soddemann, Th and D{\"u}nweg, Burkhard and Kremer, Kurt},
  journal={Eur. Phys. J. E},
  volume={6},
  pages={409--419},
  year={2001},
  publisher={Springer}
}

@Article{Santra2019,
author ="Santra, Aritra and Kumari, Kiran and Padinhateeri, Ranjith and Dünweg, B. and Prakash, J. Ravi",
title  ="Universality of the collapse transition of sticky polymers",
journal  ="Soft Matter",
year  ="2019",
volume  ="15",
issue  ="39",
pages  ="7876-7887",
publisher  ="The Royal Society of Chemistry",
doi  ="10.1039/C9SM01361J",
url  ="http://dx.doi.org/10.1039/C9SM01361J"}

@book{deGennes1979,
  title={Scaling concepts in polymer physics},
  author={De Gennes, Pierre-Gilles},
  year={1979},
  publisher={Cornell university press}
}

@article{Fixman1981,
author = {Fixman, Marshall},
title = {Inclusion of hydrodynamic interaction in polymer dynamical simulations},
journal = {Macromolecules},
volume = {14},
number = {6},
pages = {1710-1717},
year = {1981},
doi = {10.1021/ma50007a019},
URL = {https://doi.org/10.1021/ma50007a019},
}

@article{Paul1991,
  title={Crossover scaling in semidilute polymer solutions: a {M}onte {C}arlo test},
  author={Paul, Wolfgang and Binder, Kurt and Heermann, Dieter W and Kremer, Kurt},
  journal={J. Phys. II},
  volume={1},
  number={1},
  pages={37--60},
  year={1991},
  publisher={EDP Sciences}
}

@article{Prabhakar2004,
  title={A successive fine-graining scheme for predicting the rheological properties of dilute polymer solutions},
  author={Prabhakar, R and Prakash, J Ravi and Sridhar, Tamarapu},
  journal={J. Rheol.},
  volume={48},
  number={6},
  pages={1251--1278},
  year={2004},
  publisher={The Society of Rheology}
}

@article{Sunthar2005,
  title={Parameter-free prediction of DNA conformations in elongational flow by successive fine graining},
  author={Sunthar, Papanasamoorthy and Prakash, J Ravi},
  journal={Macromolecules},
  volume={38},
  number={2},
  pages={617--640},
  year={2005},
  publisher={ACS Publications}
}

@Article{Varakhedkar2025,
author ="Varakhedkar, Amit and Sunthar, P. and Prakash, J. Ravi",
title  ="Linear viscoelasticity of semiflexible polymers with hydrodynamic interactions",
journal  ="Soft Matter",
year  ="2026",
volume  ="22",
issue  ="2",
pages  ="369-386",
publisher  ="The Royal Society of Chemistry",
doi  ="10.1039/D5SM00956A",
url  ="http://dx.doi.org/10.1039/D5SM00956A"}

@article{Sahouani1992,
author = {Sahouani, H. and Lodge, T. P.},
title = {Onset of excluded-volume effects in chain dynamics},
journal = {Macromolecules},
volume = {25},
number = {21},
pages = {5632-5642},
year = {1992},
doi = {10.1021/ma00047a012},
URL = {https://doi.org/10.1021/ma00047a012},
}

@article{Clasen2006,
    author = {Clasen, C. and Plog, J. P. and Kulicke, W.-M. and Owens, M. and Macosko, C. and Scriven, L. E. and Verani, M. and McKinley, G. H.},
    title = {How dilute are dilute solutions in extensional flows?},
    journal = {J. Rheol.},
    volume = {50},
    number = {6},
    pages = {849-881},
    year = {2006},
    month = {11},
    issn = {0148-6055},
    doi = {10.1122/1.2357595},
    url = {https://doi.org/10.1122/1.2357595},
}

@article{Anderson2020,
title = {HOOMD-blue: A Python package for high-performance molecular dynamics and hard particle Monte Carlo simulations},
journal = {Comput. Mater. Sci},
volume = {173},
pages = {109363},
year = {2020},
issn = {0927-0256},
doi = {https://doi.org/10.1016/j.commatsci.2019.109363},
url = {https://www.sciencedirect.com/science/article/pii/S0927025619306627},
author = {Joshua A. Anderson and Jens Glaser and Sharon C. Glotzer}
}

@article{Fiore2017,
    author = {Fiore, Andrew M. and Balboa Usabiaga, Florencio and Donev, Aleksandar and Swan, James W.},
    title = {Rapid sampling of stochastic displacements in Brownian dynamics simulations},
    journal = {J. Chem. Phys.},
    volume = {146},
    number = {12},
    pages = {124116},
    year = {2017},
    month = {03},
    issn = {0021-9606},
    doi = {10.1063/1.4978242},
    url = {https://doi.org/10.1063/1.4978242},
}

@article{Robe2024,
author = {Robe, Dominic and Santra, Aritra and McKinley, Gareth H. and Prakash, J. Ravi},
title = {Evanescent Gels: Competition between Sticker Dynamics and Single-Chain Relaxation},
journal = {Macromolecules},
volume = {57},
number = {9},
pages = {4220-4235},
year = {2024},
doi = {10.1021/acs.macromol.3c02055},
URL = {https://doi.org/10.1021/acs.macromol.3c02055},
}

@article{Wittmer2015,
author = {J.P. Wittmer and H. Xu and O. Benzerara and J. Baschnagel},
title = {Fluctuation-dissipation relation between shear stress relaxation modulus and shear stress autocorrelation function revisited},
journal = {Mol. Phys.},
volume = {113},
number = {17-18},
pages = {2881--2893},
year = {2015},
publisher = {Taylor \& Francis},
doi = {10.1080/00268976.2015.1023225},
URL = {https://doi.org/10.1080/00268976.2015.1023225},
}

@article{Lee2009,
author = {Lee, Won Bo and Kremer, Kurt},
title = {Entangled Polymer Melts: Relation between Plateau Modulus and Stress Autocorrelation Function},
journal = {Macromolecules},
volume = {42},
number = {16},
pages = {6270-6276},
year = {2009},
doi = {10.1021/ma9008498},
URL = {https://doi.org/10.1021/ma9008498},
}

@article{Jain2012,
  title = {Dynamic {C}rossover {S}caling in {P}olymer {S}olutions},
  author = {Jain, Aashish and D\"unweg, B. and Prakash, J. Ravi},
  journal = {Phys. Rev. Lett.},
  volume = {109},
  issue = {8},
  pages = {088302},
  numpages = {4},
  year = {2012},
  month = {Aug},
  publisher = {American Physical Society},
  doi = {10.1103/PhysRevLett.109.088302},
  url = {https://link.aps.org/doi/10.1103/PhysRevLett.109.088302}
}

@Article{Pham2008,
  author  = {Tri Thanh Pham and P. Sunthar and J. Ravi Prakash},
  journal = {J. Non-Newtonian Fluid Mech.},
  title   = {An alternative to the bead-rod model: {B}ead-spring chains with successive fine graining},
  year    = {2008},
  number  = {1},
  pages   = {9-19},
  volume  = {149},
}

@article{Saadat2015,
    author = {Saadat, Amir and Khomami, Bamin},
    title = {Molecular based prediction of the extensional rheology of high molecular weight polystyrene dilute solutions: A hi-fidelity Brownian dynamics approach},
    journal = {J. Rheol.},
    volume = {59},
    number = {6},
    pages = {1507-1525},
    year = {2015},
    month = {11},
    issn = {0148-6055},
    doi = {10.1122/1.4933320},
    url = {https://doi.org/10.1122/1.4933320},
}

@incollection{LarsonBook,
title = {CHAPTER 8 - {V}iscoelasticity of {D}ilute {P}olymer {S}olutions},
editor = {Ronald G. Larson},
booktitle = {Constitutive Equations for Polymer Melts and Solutions},
publisher = {Butterworth-Heinemann},
pages = {219-270},
year = {1988},
series = {Butterworths Series in Chemical Engineering},
isbn = {978-0-409-90119-1},
doi = {https://doi.org/10.1016/B978-0-409-90119-1.50013-0},
url = {https://www.sciencedirect.com/science/article/pii/B9780409901191500130},
author = {Ronald G. Larson}
}

@article{Press1992,
  title={Numerical recipes in {F}ortran 77},
  author={Press, William H and Teukolsky, Saul A and Vetterling, William T and Flannery, Brian P},
  journal={The art of scientific computing},
  year={1992}
}

@article{Kroger2000,
	title={Variance reduced {B}rownian simulation of a bead-spring chain under steady shear flow considering hydrodynamic interaction effects},
	author={Kr{\"o}ger, Martin and Alba-P{\'e}rez, Ana and Laso, Manuel and {\"O}ttinger, Hans Christian},
	journal={J. Chem. Phys.},
	volume={113},
	number={11},
	pages={4767--4773},
	year={2000},
	publisher={American Institute of Physics}
}

@Article{Prakash1997,
  author    = {Prakash, J Ravi and {\"O}ttinger, Hans Christian},
  journal   = {J. Non-Newtonian Fluid Mech.},
  title     = {Universal viscometric functions for dilute polymer solutions},
  year      = {1997},
  number    = {3},
  pages     = {245--272},
  volume    = {71},
  publisher = {Elsevier},
}

@article{Ottinger1989,
	title={Gaussian approximation for {R}ouse chains with hydrodynamic interaction},
	author={{\"O}ttinger, Hans Christian},
	journal={J. Chem. Phys.},
	volume={90},
	number={1},
	pages={463--473},
	year={1989},
	publisher={AIP Publishing}
}

@article{Prakash2002,
	author = {Prakash, J. Ravi},
	title = {Rouse chains with excluded volume interactions in steady simple shear flow},
	journal = {J. Rheol.},
	volume = {46},
	number = {6},
	pages = {1353-1380},
	year = {2002},
	month = {11},
	issn = {0148-6055},
	doi = {10.1122/1.1514054},
	url = {https://doi.org/10.1122/1.1514054},
}

@article{Ottinger1987,
	author = {{\"O}ttinger, Hans Christian},
	title = {Generalized {Z}imm model for dilute polymer solutions under theta conditions},
	journal = {J. Chem. Phys.},
	volume = {86},
	number = {6},
	pages = {3731-3749},
	year = {1987},
	month = {03},
	issn = {0021-9606},
	doi = {10.1063/1.451975},
	url = {https://doi.org/10.1063/1.451975},
}

@Article{Ottinger1989b,
  author    = {{\"O}ttinger, Hans Christian and Rabin, Yitzhak},
  journal   = {J. {N}on-{N}ewtonian {F}luid {M}ech.},
  title     = {Renormalization-group calculation of viscometric functions based on conventional polymer kinetic theory},
  year      = {1989},
  number    = {1},
  pages     = {53--93},
  volume    = {33},
  publisher = {Elsevier},
}

@article{Sunthar2006,
	doi = {10.1209/epl/i2006-10067-y},
	url = {https://doi.org/10.1209/epl/i2006-10067-y},
	year = {2006},
	month = {jun},
	publisher = {},
	volume = {75},
	number = {1},
	pages = {77},
	author = {P. Sunthar and J. Ravi Prakash},
	title = {Dynamic scaling in dilute polymer solutions: {T}he importance of dynamic correlations},
	journal = {Europhys. Lett.}
}

@article{Prakash1999,
	title={Viscometric functions for a dilute solution of polymers in a good solvent},
	author={Prakash, J Ravi and {\"O}ttinger, Hans Christian},
	journal={Macromolecules},
	volume={32},
	number={6},
	pages={2028--2043},
	year={1999},
	publisher={ACS Publications}
}

@article{Garcia1984,
	title={Monte Carlo study of hydrodynamic properties of flexible linear chains: analysis of several approximate methods},
	author={Garcia De la Torre, Jose and Lopez Martinez, Maria C and Tirado, Maria M},
	journal={Macromolecules},
	volume={17},
	number={12},
	pages={2715--2722},
	year={1984},
	publisher={ACS Publications}
}

@article{Freire1986,
	title={Monte Carlo calculations for linear and star polymers with intramolecular interactions. 2. Nonpreaveraged study of hydrodynamic properties at the $\theta$ state},
	author={Freire, Juan J and Rey, Antonio and Garcia de la Torre, Jose},
	journal={Macromolecules},
	volume={19},
	number={2},
	pages={457--462},
	year={1986},
	publisher={ACS Publications}
}

@PhdThesis{Landry,
  author = {C. S. Landry},
  school = {University of Wisconsin},
  year   = {1985},
}

@Article{Osaki,
  author  = {Osaki, Kunihiro},
  journal = {Macromolecules},
  title   = {A Revised Version of the Intergrodifferential Equation in the Zimm Theory for Polymer Solution Dynamics},
  year    = {1972},
  number  = {2},
  pages   = {141-144},
  volume  = {5},
}

@article{Pan2014,
	title={Viscosity radius of polymers in dilute solutions: {U}niversal behavior from {DNA} rheology and {B}rownian dynamics simulations},
	author={Pan, Sharadwata and Ahirwal, Deepak and Nguyen, Duc At and Sridhar, Tamarapu and Sunthar, Papanasamoorthy and Prakash, J Ravi},
	journal={Macromolecules},
	volume={47},
	number={21},
	pages={7548--7560},
	year={2014},
	publisher={ACS Publications}
}

@incollection{Prakash1999b,
	title={The kinetic theory of dilute solutions of flexible polymers: {H}ydrodynamic interaction},
	author={Prakash, J Ravi},
	booktitle={Rheology Series},
	volume={8},
	pages={467--517},
	year={1999},
	publisher={Elsevier}
}

@article{Larson2004,
  title={An explanation for the high-frequency elastic response of dilute polymer solutions},
  author={Larson, Ronald G},
  journal={Macromolecules},
  volume={37},
  number={13},
  pages={5110--5114},
  year={2004},
  publisher={ACS Publications}
}

@article{Saha2013,
  title={Explaining the absence of high-frequency viscoelastic relaxation modes of polymers in dilute solutions},
  author={Dalal, Indranil Saha and Larson, Ronald G},
  journal={Macromolecules},
  volume={46},
  number={5},
  pages={1981--1992},
  year={2013},
  publisher={ACS Publications}
}

@article{Lodge1993,
  title={Solvent dynamics, local friction, and the viscoelastic properties of polymer solutions},
  author={Lodge, Timothy P},
  journal={J. Phys. Chem.},
  volume={97},
  number={8},
  pages={1480--1487},
  year={1993},
  publisher={ACS Publications}
}

@article{Morris1988,
  title={Solvent friction in polymer solutions and its relation to the high frequency limiting viscosity},
  author={Morris, RL and Amelar, S and Lodge, TP},
  journal={J. Chem. Phys.},
  volume={89},
  number={10},
  pages={6523--6537},
  year={1988},
  publisher={American Institute of Physics}
}

@article{Daldrop2018,
  title={Butane dihedral angle dynamics in water is dominated by internal friction},
  author={Daldrop, Jan O and Kappler, Julian and Br{\"u}nig, Florian N and Netz, Roland R},
  journal={Proc. Natl. Acad. Sci. U. S.},
  volume={115},
  number={20},
  pages={5169--5174},
  year={2018},
  publisher={National Academy of Sciences}
}

@article{Dalton2024,
  title={The role of memory-dependent friction and solvent viscosity in isomerization kinetics in viscogenic media},
  author={Dalton, Benjamin A and Kiefer, Henrik and Netz, Roland R},
  journal={Nat. Commun.},
  volume={15},
  number={1},
  pages={3761},
  year={2024},
  publisher={Nature Publishing Group UK London}
}

@article{Schulz2015,
  title={Unfolding and folding internal friction of $\beta$-hairpins is smaller than that of $\alpha$-helices},
  author={Schulz, Julius CF and Miettinen, Markus S and Netz, RR},
  journal={J. Phys. Chem. B.},
  volume={119},
  number={13},
  pages={4565--4574},
  year={2015},
  publisher={ACS Publications}
}

@article{Muller1996,
  title={Local structure and dynamics in solvent-swollen polymers},
  author={M{\"u}ller-Plathe, Florian},
  journal={Macromolecules},
  volume={29},
  number={13},
  pages={4782--4791},
  year={1996},
  publisher={ACS Publications}
}

@article{Sasmal2017,
  title={Parameter-free prediction of DNA dynamics in planar extensional flow of semidilute solutions},
  author={Sasmal, Chandi and Hsiao, Kai-Wen and Schroeder, Charles M and Ravi Prakash, J},
  journal={J. Rheol.},
  volume={61},
  number={1},
  pages={169--186},
  year={2017},
  publisher={AIP Publishing}
}

\end{document}